\documentclass[traditabstract]{aa} 
\usepackage{graphicx}
\usepackage{amsmath}    
\usepackage{verbatim}   
\usepackage{color}      
\usepackage{subfigure}  
\usepackage{hyperref}   
\usepackage{natbib}
\usepackage{multirow} 
\usepackage{txfonts}
\begin{document}

   \title{Infrared imaging and polarimetric observations \\
   of the pulsar wind nebula in SNR~G21.5-0.9}


   \author{A. Zajczyk
          \inst{1, 2}\fnmsep\thanks{zajczyk@ncac.torun.pl}
          \and
          Y.A. Gallant\inst{2}
		  \and
		  P. Slane\inst{3}
		  \and
		  S.P. Reynolds\inst{4}
		  \and				  
		  R. Bandiera\inst{5}
		  \and
		  C. Gouiff\`{e}s\inst{6}
		  \and
		  E. Le Floc'h\inst{6}
		  \and 
		  F. Comer\'{o}n\inst{7} 
		  \and
		  L.~Koch Miramond\inst{6}       
          }

   \institute{Department of Astrophysics, N. Copernicus Astronomical Center,       
   			  ul. Rabia\'{n}ska 8, 87-100 Toru\'{n}, Poland
         	\and
              LUPM, Universit\'e Montpellier 2, CNRS/IN2P3,
              place E. Bataillon, 34095 Montpellier, France
         	\and
			  Center for Astrophysics (CfA), 60 Garden Street,
			  Cambridge, MA 02138, USA
			\and
			  North Carolina State University Physics Dept.,
			  Box 8202, Raleigh, NC 27695-8202, USA		
			\and
			  INAF - Osservatorio Astrofisico di Arcetri, 
			  Largo E. Fermi 5, I-50125 Firenze, Italy
			\and
			  Service d'Astrophysique, CEA-Saclay,
			  91191 Gif-sur-Yvette Cedex, France 
			\and
              ESO, Karl-Schwarzschild-Str. 2, 85748 Garching 
              bei M\"{u}nchen, Germany
             }

   \date{Received ... 2011; accepted ...}

   
  \abstract 
  {We present infrared observations of the supernova remnant G21.5-0.9 with the Very Large 
   Telescope, the Canada-France-Hawaii Telescope and the \textit{Spitzer} Space 
   Telescope. Using the VLT/ISAAC camera equipped with a narrow-band [Fe~II]
   1.64~$\mu$m filter the entire pulsar wind nebula in SNR G21.5-0.9 was imaged. 
   This led to detection of iron line-emitting material in the shape of 
   a broken ring-like structure following the nebula's edge. The detected
   emission is limb-brightened. We also detect the compact nebula
   surrounding 
   PSR~J1833-1034, both through imaging with the CFHT/AOB-KIR instrument
   ($K'$ band) and the IRAC camera (all bands) and also 
   through polarimetric observations performed with VLT/ISAAC ($Ks$ band). 
   The emission from the compact nebula is highly polarised with an average
   value of the linear polarisation fraction $P_{\mathrm{L}}^{avg} \simeq 0.47$,
   and the swing of the electric vector across the nebula can be observed.
   The infrared spectrum of the compact nebula can be described as a
   power law of index $\alpha_{\rm IR} = 0.7 \pm 0.3$, and suggests that
   the spectrum flattens between the infrared and X-ray bands.
   }


   \titlerunning{IR imaging and polarimetric observations of the PWN in SNR G21.5-0.9}
   \authorrunning{A. Zajczyk et al.}

   \maketitle

\section{Introduction}
\label{intro}

In the past years it has been shown that the infrared band is an important
energy range for studying supernova remnants (SNRs) \citep[e.g.,][]{oliva89,
koo07,lee09,tm10} and pulsar wind nebulae (PWNe) \citep[e.g.,][]{grah89,tm06,sln08,wil08}, 
especially in the cases where the object is heavily obscured by dust 
in the optical band. 
Infrared observations allow for detecting synchrotron emission from relativistic
particles pervading the PWN, e.g. the Crab Nebula \citep{tm06}, 3C~58 \citep{sln08},
B0540-69.3 \citep{wil08}, 
and line-emitting material excited either by a passing shock, e.g. G11.2-0.3
\citep{koo07}, 3C~396 \citep{lee09}, RCW~103 \citep{oliva89} or 
by photoionization by the synchrotron continuum, e.g. filaments in the Crab 
Nebula \citep{grah89,tm06}. This in turn provides essential 
information about the dynamical evolution and physical conditions 
in these objects.
Infrared emission may also include a component of dust continuum, 
giving information on dust in the remnant vicinity or freshly synthesized 
in the ejecta \citep{wil08}.
Moreover, polarisation studies in the optical or near infrared
may be an effective tool for studying the inner regions of PWNe as proposed
by recent MHD simulations \citep{bcc05,delz06}. Through polarimetry we can
learn about the magnetic field structure and the flow speed in the wind 
termination shock area. 

The supernova remnant G21.5-0.9 belongs to the class of composite 
SNRs, characterised by a central pulsar-powered component, accompanied
by an expanding shell of material swept up by the supernova blast wave.
Extensive studies of G21.5-0.9 carried out in the radio \citep[e.g.][]{beck81, fur88, bb08}
and in X-rays \citep[e.g.][]{sln00, sfh01, bocc05, ms10} reveal the complex structure 
of this object. It consists of a radio and X-ray bright pulsar wind nebula of 
radius $\sim 40''$ which is 
surrounded by a diffuse and faint X-ray halo, due to dust scattering, with
a brighter limb of radius $\sim 150''$, which traces the shell of the associated
SNR \citep{bocc05}. A very X-ray bright compact source (radius $\sim 2''$) 
is situated in the centre of the system. It harbours one 
of the most energetic pulsars in our Galaxy with a spin-down luminosity 
of $\dot{E} = 3.3 \times 10^{37}$ erg s$^{-1}$ \citep{gpt05,cam06}.
The pulsar J1833-1034 has a period of 61.8 ms and a 
characteristic age of 4800 years. A recent age estimate based on a
measurement of the PWN expansion rate in the radio band \citep{bb08}
places this pulsar among the youngest in our Galaxy, with an age of
$870^{+200}_{-150}$~yr (or even lower, in the case of decelerated
expansion), much less than the characteristic spin-down time.
The distance to the system is estimated, based on HI and $^{13}$CO measurements, 
to be 4.8 kpc \citep{tian08} with an uncertainty of 0.4 kpc (Tian, personal communication).
Thus, SNR~G21.5-0.9 is one of only a few ``complete'' systems in which 
a shell supernova remnant is observed to contain a pulsar 
and pulsar-wind nebula. In some composite SNRs, the pulsar is not directly 
observed; in some young PWNe, a shell is either not observed or can be separated 
from the PWN only with difficulty. An object very similar to G21.5-0.9 
is B0540-69.3, with a similar age and pulsar spindown luminosity, but its location 
in the LMC (50 kpc distant) makes spatial resolution of the PWN much more difficult.

Due to its position in the inner Galaxy, G21.5-0.9 is heavily obscured in 
the optical band, where the estimated value of interstellar extinction 
derived from the hydrogen column density $N_{\mathrm{H}} \simeq 2.2 \times 
10^{22}$ cm$^{-2}$ \citep{sln00} reaches $A_{\mathrm{V}} \simeq 10$
\citep{gor75}.  The interstellar extinction drops dramatically when
moving to longer wavelengths \citep{crdl89},
yielding an estimated near-infrared extinctions $A_K \simeq 1.2$ and 
$A_H \simeq 1.9$ toward G21.5-0.9.
The first mid-infrared observations of the remnant taken with the
ISOCAM instrument on-board the ISO observatory show extended emission
which matches the morphology of the PWN, most clearly at 15\,$\mu$m
\citep{galtuf99}.  If interpreted as synchrotron emission, the
extracted infrared fluxes fitted together with the previously
existing radio \citep[][and references therein]{salt89} and X-ray
data \citep{dav86, ak90} by a broken power-law model
would imply a spectral index of $\alpha_{\mathrm{X}} \simeq 1.0$
($F_{\nu} \propto E^{- \alpha_{\mathrm{X}}}$)
between the infrared and X-ray fluxes. 

SNR G21.5-0.9 was observed in the near-infrared band with the Canada-France-Hawaii 
Telescope, the ESO Very Large Telescope and the \textit{Spitzer} Space
Telescope with the aim 
of detecting synchrotron emission originating from relativistic 
particles accelerating in the PWN magnetic field, and detecting clumped 
supernova ejecta and/or filaments. A detailed description of the
observations and data analysis can be found in Sect.~\ref{obs}. We present 
results in Sect.~\ref{res}, their interpretation in Sect.~\ref{interp}
and conclusions in Sect.~\ref{concl}.


\section{Observations and data analysis}
\label{obs}

\subsection{CFHT/AOB-KIR}
\label{sub-cfht}
Imaging of G21.5-0.9 was carried out using the Canada-France-
Hawaii Telescope equipped with the Adaptive Optics Bonnette (AOB) and the
KIR near-infrared camera (together named AOB-KIR\footnote{http://www.
cfht.hawaii.edu/Instruments/Detectors/IR/KIR/}), in order both to look
for fine structure in the synchrotron nebula and to mitigate confusion
with stars in this crowded Galactic field.  The observations were 
performed in the \textit{K'} filter, a slightly shorter-wavelength
version of the standard \textit{K} filter designed to reduce
atmospheric background. The details of the 
obtained data can be found in Table~\ref{cfht-data}. Basic data 
reduction was carried out using the \textit{noao.imred.ccdred} package of the IRAF\footnote{IRAF 
is distributed by the National Optical Astronomy Observatory, which is operated by 
the Association of Universities for Research in Astronomy (AURA) under cooperative 
agreement with the National Science Foundation.} Data Reduction and Analysis Facility.
For each individual exposure, bias and dark were subtracted, the flat-field correction was 
applied using dome flats and bad pixels were suppressed.

\begin{table*}
\begin{minipage}[t]{\columnwidth}
\caption{The CFHT/AOB-KIR and VLT/ISAAC observations of G21.5-0.9}
	\centering
	\renewcommand{\footnoterule}{}  
	\begin{tabular}{lccccc}
	\hline\hline
	Instrument	& Number	& Exposure 		& Filter		& Filter	& Date\\
	name		& of		& time per frame	& central $\lambda$	& width	$\Delta\lambda$	&of\\ 
			& exposures	& [s]			& [$\mu$m]		& [$\mu$m]		&observations\\\hline
	\multicolumn{6}{c}{CFHT}\\\hline
	AOB-KIR	& 31		& 120.0 		& 2.115		& 0.350	& 02.08.2001\\
	AOB-KIR	& 54		& 120.0 		& 2.115		& 0.350	& 03.08.2001\\
	&&&&&\\\hline
	\multicolumn{6}{c}{VLT}\\\hline
	ISAAC-\textit{POL}\tablefootmark{a}  & 		& 		& 		& 	&\\
	$\chi = 27\degr$ & 9	& 20.0 		& 2.160		& 0.270	& 07.09.2002\\
		& 27	& 20.0 & 2.160 & 0.270	& 08.09.2002\\
	$\chi = 72\degr$ & 2	& 20.0 		& 2.160		& 0.270	& 07.09.2002\\
		& 34	& 20.0 & 2.160 & 0.270	& 08.09.2002\\
	&&&&&\\
	ISAAC-\textit{IMG}\tablefootmark{b}  & 		& 		& 		& 	&\\
		& 9	& 68.0	& 1.64	& 0.025	& 08.07.2002\\
		& 9	& 68.0	& 1.64	& 0.025	& 22.07.2002\\
		& 10	& 68.0	& 1.71	& 0.026	& 08.07.2002\\
		& 10	& 68.0	& 1.71	& 0.026	& 22.07.2002\\
	\hline
	\end{tabular}
	\tablefoot{
	\tablefoottext{a}{Polarisation mode of ISAAC instrument (Sect.~\ref{sub-pol})}
	\tablefoottext{b}{Imaging mode of ISAAC instrument (Sect.~\ref{sub-img})}
	}
	\label{cfht-data}
\end{minipage}
\end{table*}

\begin{figure}
  \begin{center}
	\includegraphics[scale=0.45,angle=-90]{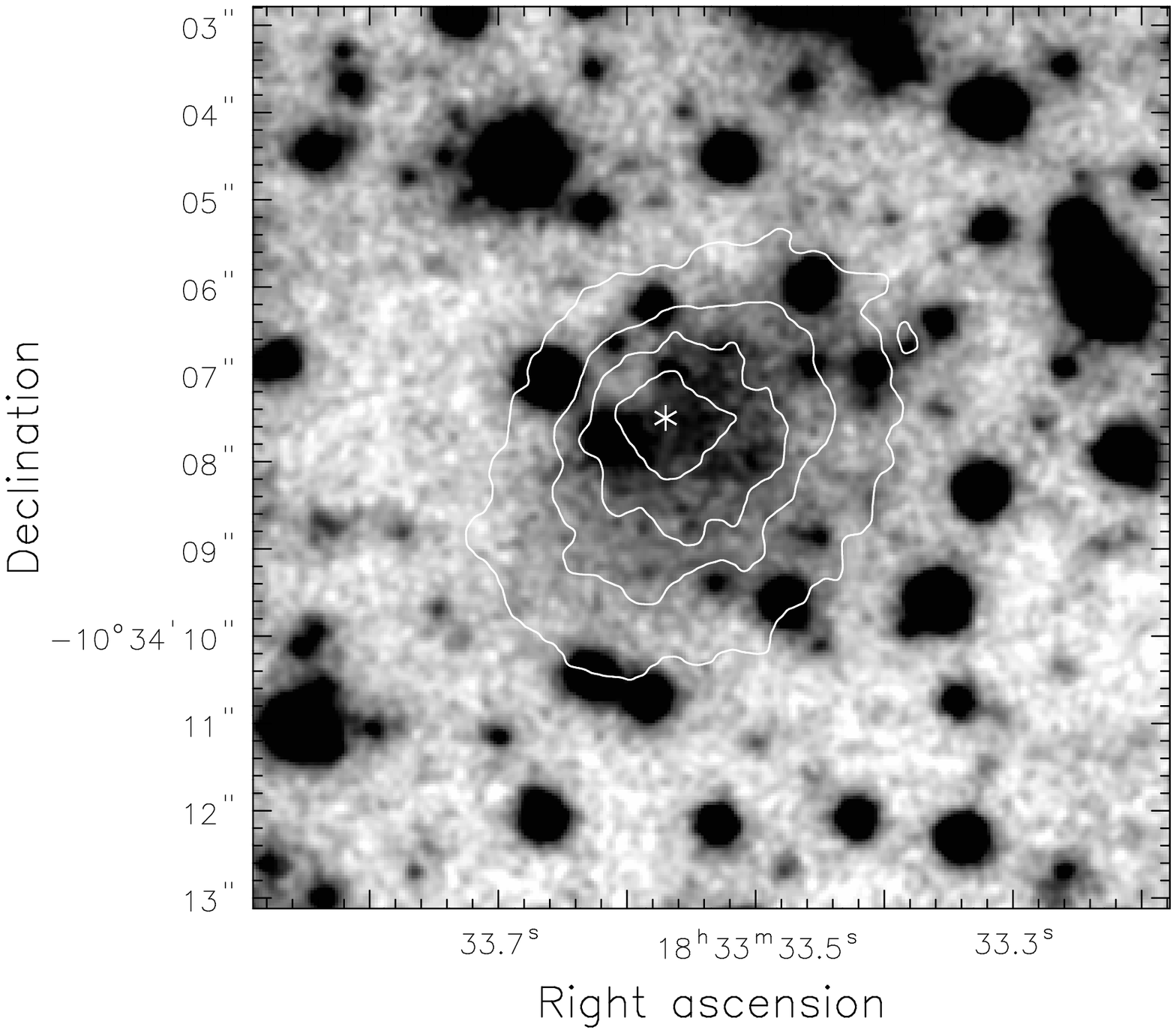}\\
	
	\vspace{0.3cm}\includegraphics[scale=0.45,angle=-90]{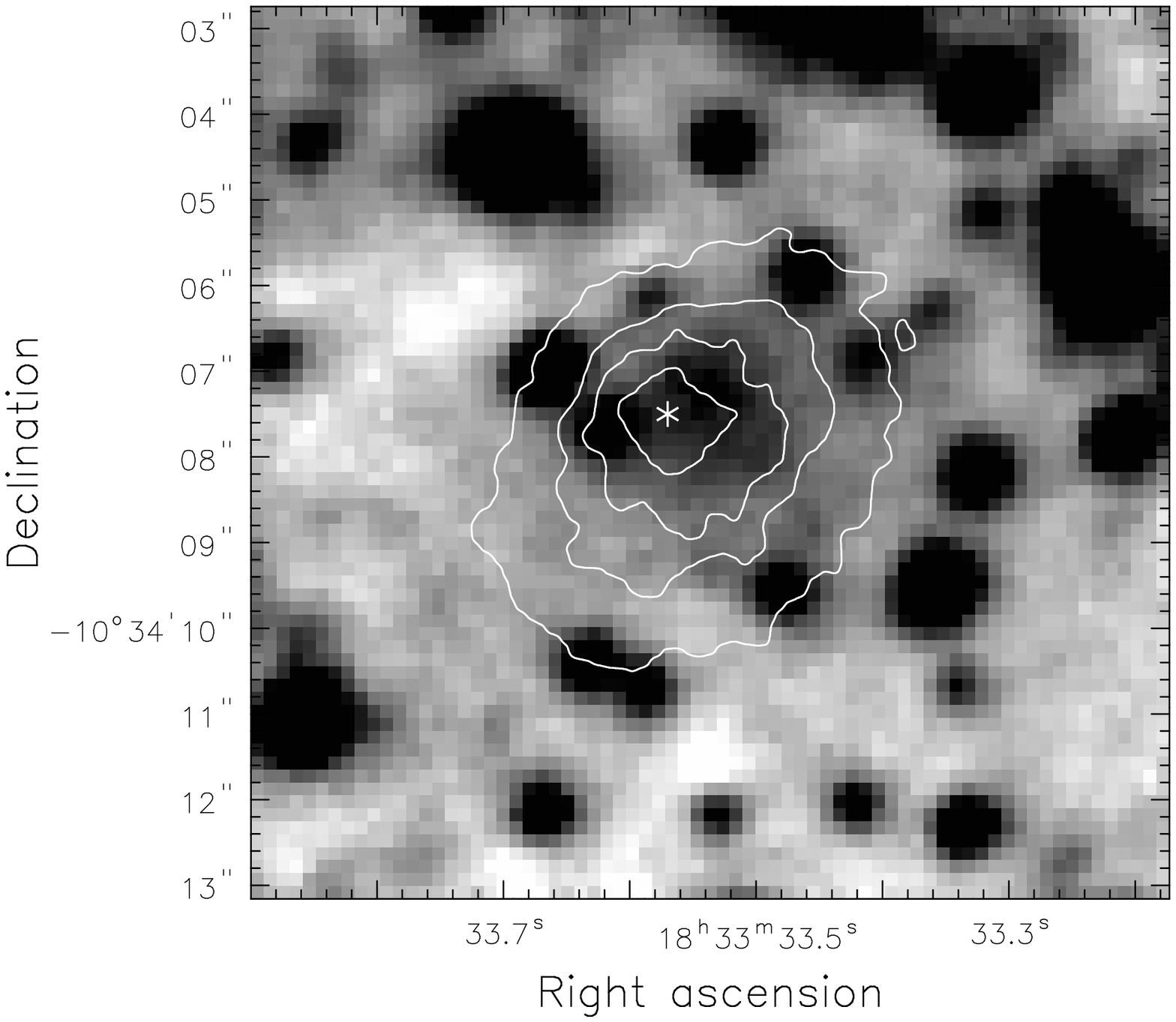}

	\caption{\small \textit{Top panel:} CFHT/AOB-KIR mosaicked image of the central 
	part of SNR G21.5-0.9 taken in \textit{K'} band, smoothed with 2-dimensional Gaussian 
	with a $FWHM = 0.11''$. \textit{Bottom panel:} the same sky region as in 
	the \textit{top} panel imaged with VLT/ISAAC in \textit{Ks} band. The 
	image shown is the averaged total intensity $I$ (see Sect.~\ref{sub-pol}).
	For both panels a star depicts the position of PSR~J1833-1034, while the extent 
	of the compact core seen with {\it Chandra} HRC \citep{cam06} is shown with contours.
	The observations taken with CFHT/AOB-KIR have higher resolution than the ones obtained 
	with VLT/ISAAC, which has direct implications for a flux determination of an extended 
	emission structure visible around the pulsar position (see Sect.~\ref{res}).}
	\label{cfht-img}
  \end{center}
\end{figure}

Due to the small field of view of the instrument 
($36\arcsec \times 36\arcsec$) as compared to the 
expected size of the source, successive observations were taken following 
a raster pattern on the sky so as to cover the desired field. In order 
to reconstruct the full image of the sky covered by the observations a 
mosaicking procedure using the IDL programming language was developed. 
The procedure makes use of the IDL Astronomy User's Library functions 
\textit{CORREL\_IMAGES} and \textit{CORRMAT\_ANALYZE} to determine 
relative shifts between subimages with 1 pixel accuracy.  A sky level 
for each of the subexposures is first computed and  
subtracted using IDL DAOPHOT \textit{SKY} procedure. 
(This has the consequence that any large-scale emission extending
smoothly over most of the $36'' \times 36''$ subexposure
field of view would be subtracted out; any such large-scale synchrotron
emission could however be detectable in polarisation with VLT/ISAAC,
see Sect.~\ref{sub-pol}.)
Having determined the offsets, in the final step a 
mosaic (size of $\sim 50\arcsec \times 60\arcsec$)
is created by stacking up and averaging the offset and sky 
corrected subimages. The \textit{top} panel of Fig.~\ref{cfht-img} shows 
the final mosaicked image. The presented region is smaller
than the size of the PWN in SNR G21.5-0.9 and covers the X-ray compact nebula region
\citep[e.g.][]{cam06}.

Within a circular region of $15''$ in diameter centred on the pulsar 
position, stars which have their brightness in the \textit{2MASS} \textit{K} band 
\citep{2mass} determined were identified. Their fluxes, measured via aperture 
photometry, allowed for calculation of a photometric zero point between 
the instrumental system and the \textit{2MASS K} band. The photometric 
zero point was later used in the transformation of the measured flux to 
an apparent magnitude for the extended emission structure detected around 
the pulsar position.

\subsection{VLT/ISAAC - polarimetry}
\label{sub-pol}
Polarimetric observations of G21.5-0.9 were taken in the \textit{Ks} 
filter with the ISAAC\footnote{http://www.eso.org/sci/facilities/
paranal/instruments/isaac/} instrument mounted on the ESO Very Large 
Telescope. The instrument uses a Wollaston prism, which splits the incoming 
radiation beam into two perpendicularly polarised beams, giving on 
one CCD plate two images of the same object separated by $\sim 21''$. In 
order to avoid sources overlapping, a special mask made of alternating 
opaque (width $\sim 24''$) and transmitting stripes (width $\sim 20''$) 
is inserted before the Wollaston prism.
This instrumental set-up produces a CCD 
image containing altogether six $20\arcsec \times 150\arcsec$ subimages of the sky - 
three different stripes of the sky imaged at the same time in two 
perpendicular polarisations.  Due to the non-standard mosaic
reconstruction required, dedicated procedures performing the basic 
data reduction were developed. A sky flat field 
was obtained for each of the six stripes separately from median 
combination of all the images taken with the same angle $\chi$ (see below), 
each normalised to its median.
A raster pattern was applied when performing the 
polarimetric imaging of G21.5-0.9 in order to obtain continuous
coverage of a FOV containing the entire PWN.
The complete image in any of the polarisation beams 
was then obtained using the mosaicking procedure described in Sect.~\ref{sub-cfht},
adapted to the specific character of the ISAAC polarimetric data, including
sky level determination for each stripe in each image.

The polarimetric imaging was carried out with two different position 
angle settings of the instrument $\chi$ (measured North to East), namely
with $\chi = 27\degr$ and $\chi = 72\degr$, allowing for 
determination of a linear polarisation degree $P_{\mathrm{L}}$, which in 
terms of the Stokes parameters $I$, $Q$ and $U$ \citep[see e.g.][for 
the Stokes parameters definition]{sperel97}
can be expressed as:
\begin{equation}
P_{\mathrm{L}} = \sqrt{P_{\mathrm{Q}}^{2} + P_{\mathrm{U}}^{2}} \, ,
\label{poldeg}
\end{equation}
where:
\begin{equation}
P_{\mathrm{Q}} = \frac{Q}{I} ~~\mathrm{and}~~ P_{\mathrm{U}} = \frac{U}{I} \,
\end{equation}
and determination of a polarisation angle:
\begin{equation}
\theta = 0.5 ~\mathrm{arctan}~\left(\frac{P_{\mathrm{U}}}{P_{\mathrm{Q}}}\right) ~~.
\label{polang}
\end{equation}
We note that the instrumental set-up is unable to measure any circular
polarisation which might also be present.
The derived polarisation angle (Eq.~\ref{polang}) is expressed in the detector 
reference frame. 
To refer it to the celestial reference frame a counter-rotation 
of $27\degr$ with respect to the instrument position angle is applied 
following e.g.\ Eq. (9) of \citet{landi07}.
In the celestial reference frame the polarisation angle is 
measured with respect to the North Celestial Meridian (North to East).
The \textit{bottom} panel of Fig.~\ref{cfht-img} shows the central part 
of G21.5-0.9, where the X-ray compact nebula is situated, as seen in 
the total intensity $I$.

Polarisation standard stars were also observed and their linear polarisation 
degree and angle were measured in order to verify determination of 
polarisation parameters for the compact nebula region. The standard 
stars were chosen following the ISAAC list of polarised standard stars. 
We used R CrA15 and HD 188112. The measurements for R CrA15 in the \textit
{Ks} filter yield $P_{\mathrm{L}} = 0.017 \pm 0.004$ and $\theta = 95^
{\circ} \pm 13\degr$ determined North through East. 
The result is in fairly good agreement with the known polarisation for this object at
shorter wavelengths - e.g., in the \textit{H} band $P_{\mathrm{L}, H} = 0.0144 \pm 0.0006$ 
and $\theta_{H} = 100\degr \pm 2\degr$ and in the \textit{J} band
$P_{\mathrm{L}, J} = 0.0228 \pm 0.0003$ and $\theta_{J} = 96\degr \pm 1\degr$ 
\citep{whittet92}.
For HD 188112 we get $P_{\mathrm{L}} 
= 0.007 \pm 0.005$ and $\theta = 118\degr \pm 14\degr$.

\begin{figure}
  \begin{center}
	\includegraphics[scale=0.35,angle=-90]{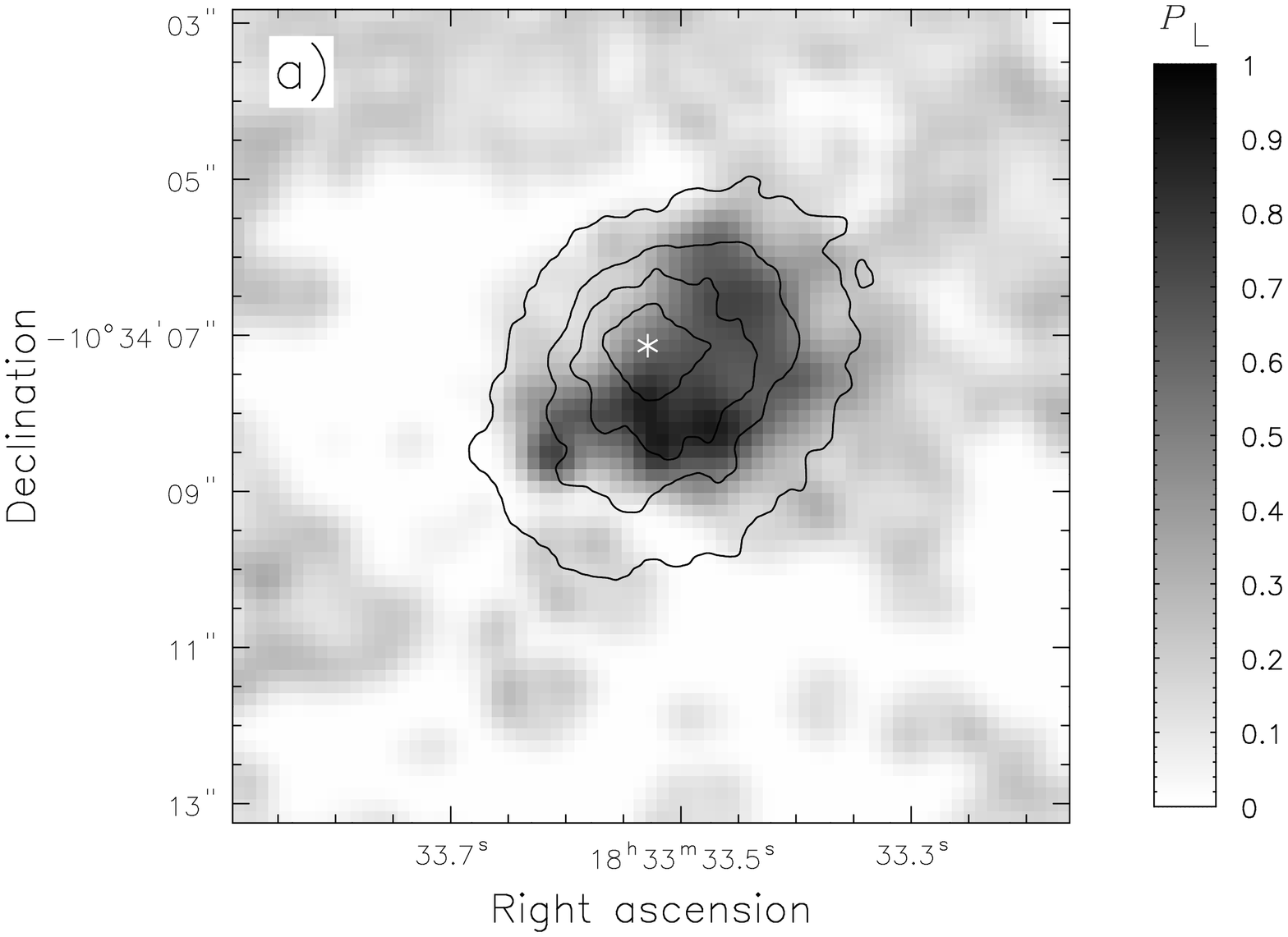}\\
	
	\vspace{-1.1cm}\includegraphics[scale=0.35,angle=-90]{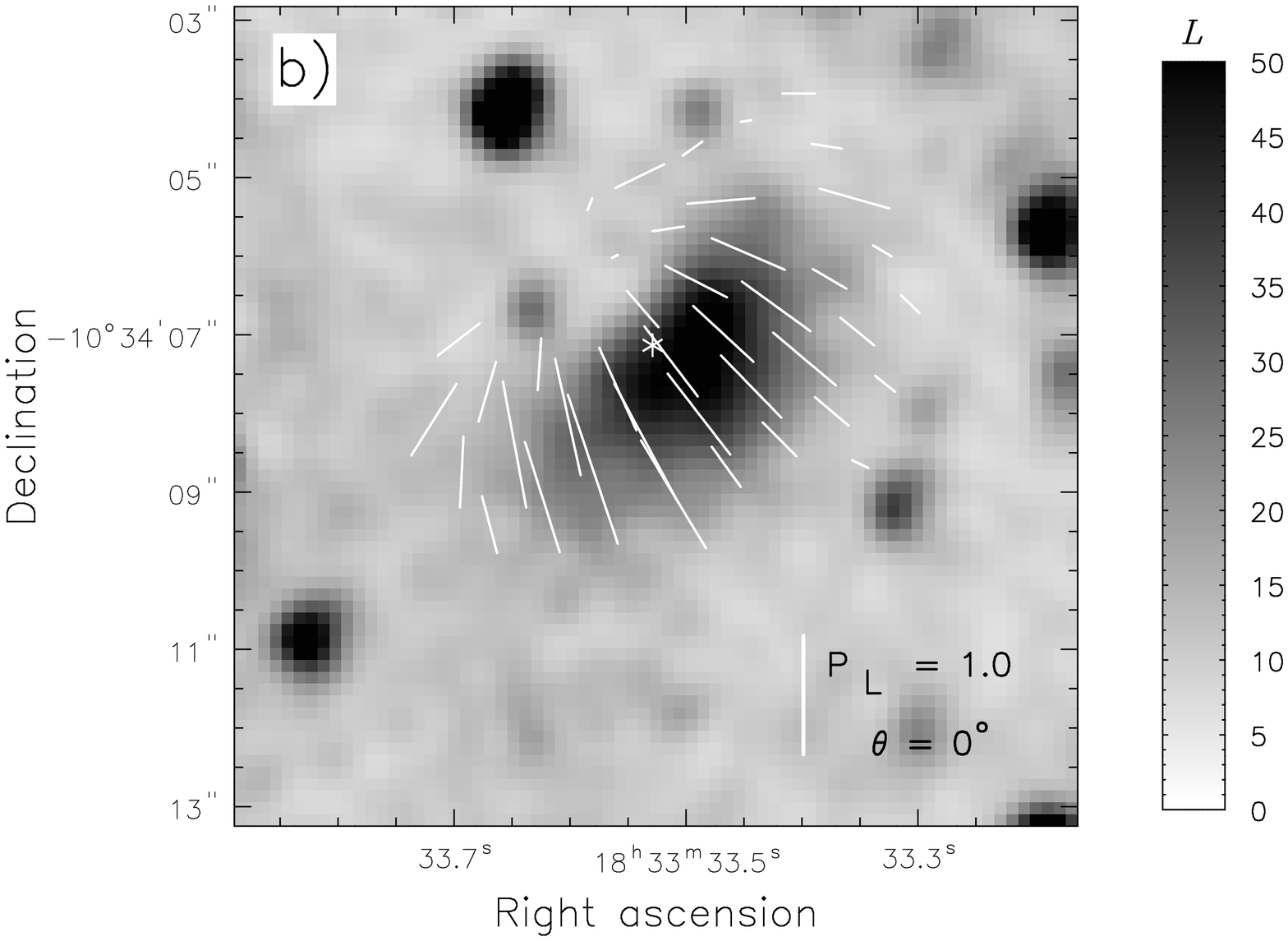}
	\caption{\small \textbf{a)} Linear polarisation degree $P_{\mathrm
	{L}}$ map of the compact core region of SNR G21.5-0.9. The map was 
	constructed using polarimetric imaging observations performed with 
	VLT/ISAAC and was smoothed using a 2-dimensional Gaussian with $FWHM = 0.45''$.
	Colour-coded is the measured value of $P_{\mathrm{L}}$. Contours show 
	the extent of the compact core seen with {\it Chandra} HRC \citep{cam06}. 
	\textbf{b)} Map of linearly polarised intensity $L = \sqrt{Q^{2}+U^
	{2}}$ smoothed with a 2-dimensional Gaussian with a $FWHM=0.45''$. Colour-coded 
	is the measured value of $L$. White bars represent electric field vectors 
	with their length proportional to the linear polarisation fraction $P_
	{\mathrm{L}}$ and their orientation defined by the polarisation angle $
	\theta$. A vertical bar placed in the bottom right corner, for reference, 
	shows an electric field vector with $P_{\mathrm{L}} = 1$ and $\theta = 
	0\degr$.	
	 In both panels a star shows the pulsar's position.}
	\label{isaac-pol-vect}
  \end{center}
\end{figure}

For the central region of G21.5-0.9 two maps, of linear polarisation 
fraction $P_{\mathrm{L}}$ and linear polarisation intensity 
$L = \sqrt{Q^{2}+U^{2}}$, were created. They are presented 
in Figure~\ref{isaac-pol-vect}. For presentation purposes pixels with 
signal of $1\sigma$ below the sky value - as determined in the total 
intensity image $I$ - were masked in the $P_{\mathrm{L}}$ map. In 
both maps an extended blob of polarised emission surrounding the 
PSR~J1833-1034 position is visible. In order to show the behaviour
of the polarisation angle across the nebula the electric 
field vectors are overlaid on the $L$ map. They were determined 
from the signal integrated over circular regions ($r \sim 0.45''$) 
spread across the polarised emission nebula. A gradual change in 
the polarisation angle is visible when moving from the SE to NW
part of the torus region and its implications are discussed in 
more detail in Sect.~\ref{wts}.

Similarly as for the CFHT observations, the flux of the emission 
coincident with the putative pulsar wind torus identified in the averaged 
total intensity image $I$ was determined. To compute the photometric 
zero point, flux measurements for the polarimetric standards were 
used together with their \textit{2MASS K} magnitudes.

\subsection{Spitzer}
\label{spitzer-img}
\begin{figure}
  \begin{center}
	\begin{tabular}{cc}	
	\hspace{-0.3cm}\includegraphics[scale=0.24,angle=-90]{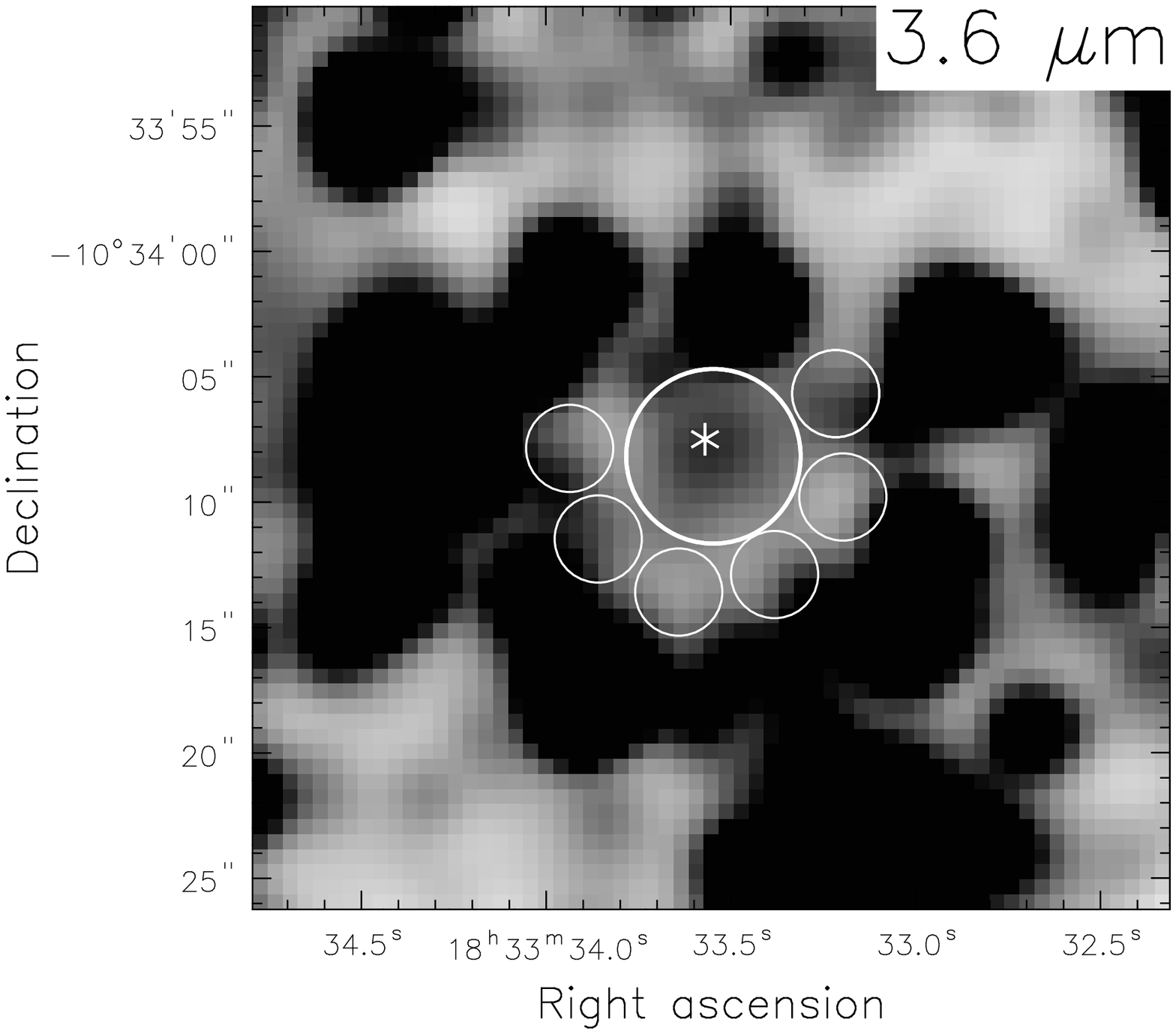}&
	\hspace{-0.3cm}\includegraphics[scale=0.24,angle=-90]{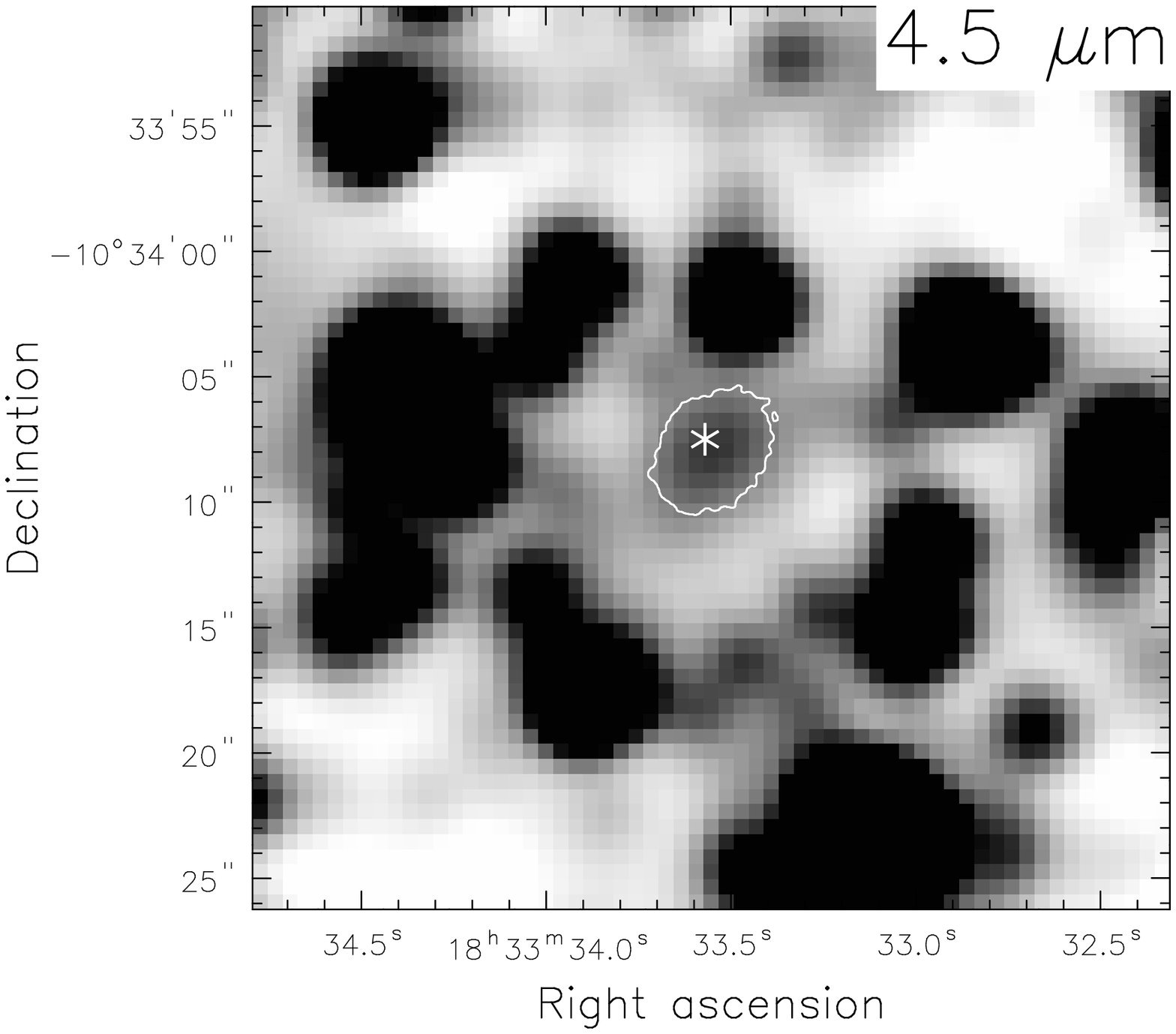}\\
	\hspace{-0.3cm}\includegraphics[scale=0.24,angle=-90]{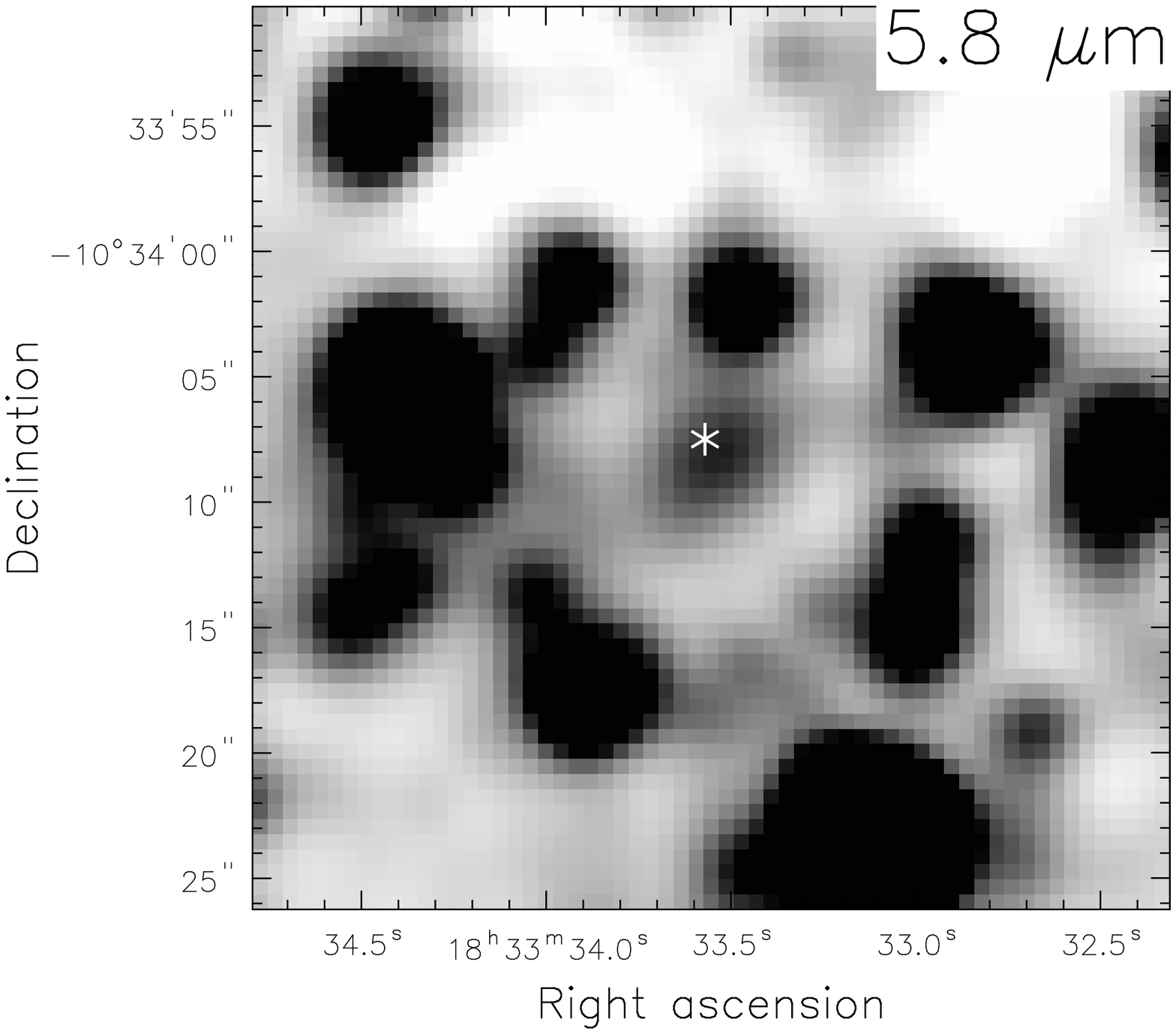}&
	\hspace{-0.3cm}\includegraphics[scale=0.24,angle=-90]{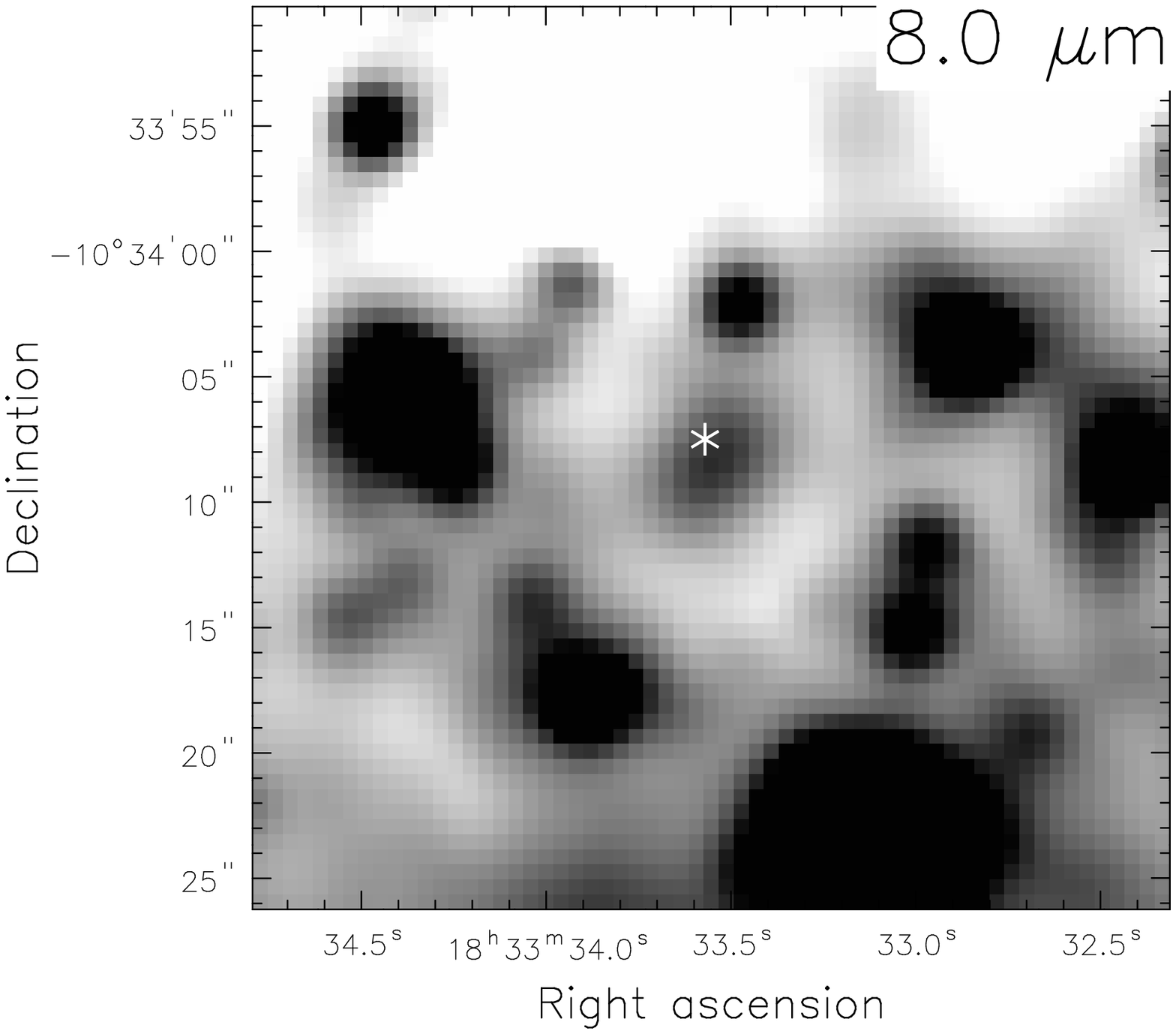}\\
	\end{tabular}
	\caption{\small IRAC images of the central region of SNR G21.5-0.9: 3.6~$\mu$m - \textit{top left},
	 4.5~$\mu$m - \textit{top right}, 5.8~$\mu$m - \textit{bottom left} and 8.0~$\mu$m - \textit{bottom right}.
	All panels cover the same field of view. Contours in the 4.5~$\mu$m image show the extent of the compact 
	nebula as seen with {\it Chandra} HRC \citep{cam06}. The star marks the position of PSR~J1833-1034.
	In the \textit{top left} panel circles depict: large central circle - flux extraction region for 
	the compact nebula; smaller circles: regions where the background level was estimated. These
	regions were used for measuring the flux in all IRAC bands (see Sect.~\ref{sub-sptz}). 
	Each image was smoothed with a 2-dimensional Gaussian with $FWHM = 1.8''$.} 
	\label{sptz-f1}
  \end{center}
\end{figure}

SNR G21.5-0.9 was observed with the Infrared Array Camera (IRAC) 
and Multiband Imaging Photometer (MIPS) onboard the \textit{Spitzer} Space
Telescope as part of observing programme ID~3647. 
Using the \textit{Spitzer} Heritage Archive, \textit{post}-Basic Calibrated 
Data (\textit{post}-BCDs) containing mosaicked images of the G21.5-0.9 region 
were obtained for all IRAC bands and the 24~$\mu$m MIPS band.
The IRAC observations at 3.6, 4.5, 5.8 and 8.0~$\mu$m were carried out on 
15 September 2005, while the MIPS 24~$\mu$m observation was taken on 11 April 2005.
Figure~\ref{sptz-f1} shows the central part of G21.5-0.9 in all IRAC
bands.

\subsection{VLT/ISAAC - [Fe~II] imaging}
\label{sub-img}
G21.5-0.9 was also observed in the [Fe~II] 1.64~$\mu$m filter and in 
a neighbouring narrow-band 1.71~$\mu$m filter with the ISAAC instrument 
(in standard imaging mode). 
A dark correction was applied to all frames following 
the standard approach using IDL procedures written to handle the ISAAC data.
In each of the filters the successive exposures were performed in a wide raster
pattern so as to cover the sky area where emission from the PWN, as well as a dense
part of the SNR shell, could be expected.  Large shifts between exposures were used
to ensure that a given camera pixel fell outside any diffuse emission from G21.5-0.9
in most frames.  A sky flat field for each of the narrow-band filters could
then be constructed by median combination of all exposures taken on the same
night (each scaled such that its median was unity).  These sky flat fields
were divided into the science frames following the standard approach. 
The IDL mosaicking procedure that also determines the sky level for each of 
the subexposures (see Sect.~\ref{sub-cfht}) was applied to obtain the final image 
in both the [Fe~II] 1.64~$\mu$m and 1.71~$\mu$m filters.

In order to minimise confusion from faint stars which is evident in
the [Fe~II] 1.64~$\mu$m image (Fig.~\ref{isaac-img-fe2}, panel~\textit{a}), we
subtracted from it the 1.71~$\mu$m image (Fig.~\ref{isaac-img-fe2}, panel~\textit{b}).
To compensate for the difference in stellar fluxes in different filters, 
an overall normalisation factor was computed from measured star 
fluxes in both narrow-band filters ($F_{1.64\mu m}/F_{1.71\mu m} \sim 0.84$)
assuming that star colour effects can be neglected over the small wavelength difference.
A simple subtraction of the aligned and scaled 1.71~$\mu$m image yields rather
conspicuous stellar residua, however, due in large part to different PSFs in
the two images owing to changing observing conditions.
An improved subtraction was thus performed using a PSF equalisation procedure,
which also ensured image alignment to sub-pixel accuracy, proceeding in the
following way.  We used 20 copies of the two original images with integer
pixel offsets (all those with a total 2-D offset less than 2.5 pixels),
and computed the linear combination of all these images, and a uniform
background, which yielded minimal residua and a minimal contribution from
the offset images.  This optimisation was performed using a central image
area which had negligible diffuse emission but a PSF as close as possible
to the average one within the PWN area. 
The result is shown in panel \textit{c} of Fig.~\ref{isaac-img-fe2}: the
effectiveness of the subtraction procedure, in the PWN area in particular,
can be gauged by comparison with panel \textit{a}, which shows the
unsubtracted image in the same intensity scale.
Some residua remain, including ''halos'' around brighter stars, and may be due
in part to non-linearities in instrument response, as well as to intrinsic
differences in stellar spectra between the 1.64~$\mu$m and 1.71~$\mu$m bands.

Panel \textit{c} of Figure~\ref{isaac-img-fe2} reveals,
after the above subtraction, diffuse emission in the [Fe~II] 1.64~$\mu$m
filter coincident with parts of the PWN in G21.5-0.9.
Extended emission in the shape
of a partial ring is clearly visible. {\it Chandra} ACIS contours are overlaid
to show the PWN extent in X-rays.
The surface brightness for the NE and S rims of the [Fe~II]
1.64~$\mu$m emission was computed. 
The flux calibration was performed using stars that have 
\textit{2MASS} \textit{H} magnitudes determined and that could be 
identified in the observed field. Detailed discussion of the results can 
be found in Sect.~\ref{sub-pwn}.

\section{Results}
\label{res}

\subsection{Compact Core -- near-infrared}
\label{sub-cc}
The field of view (\textit{FOV}) of the central part of the AOB-KIR
mosaicked image with the best S/N
is approximately $36'' \times 36''$. It covers the central part of the 
PWN as identified by X-ray observations, and in particular the compact
nebula around the pulsar \citep{sln00}. The observed 
field is rich in stars which can confuse any NIR emission coming from 
G21.5-0.9. Using a brightness scale in which stars are saturated, a
clearly extended region of emission (compared
to the instrument's $FWHM_{\mathrm{AOB-KIR}} \simeq 0.24''$) emerges from 
the background (Fig.~\ref{cfht-img}, \textit{top} panel). 
Its size of $\sim 4''$ in diameter is 
comparable with the size of the X-ray compact nebula. Its apparent 
centre is slightly shifted towards the SW with respect to the pulsar 
position $\alpha = 18^{\mathrm h} 33^{\mathrm m} 33\fs57$ and 
$\delta = - 10\degr 34' 07\farcs5$ \citep{cam06},
but more extended emission may be confused with the stars that are
situated to the NE of the pulsar. 
It is worth noting that the X-ray morphology of the emission region
around the pulsar position \citep{cam06, ms10} is similar 
to the one seen in the present near-infrared data, and can 
be caused by a Doppler-boosted toroidal structure surrounding 
the pulsar.

To extract the flux from the extended emission nebula a 
circular aperture $6''$ in diameter around the pulsar position was used.
The contributions to the flux of the nebula from stars located within
a $5''$ radius of the pulsar were subtracted by performing aperture
photometry.
The measured flux for the compact nebula corrected for interstellar 
extinction is $F_{K'} = (0.82 \pm 0.60)$~mJy. 
The existence of this emission blob is confirmed by the polarimetric 
observations. No other significant diffuse emission is detected.

Each of the mosaicked images, \textit{i(0$\degr$), i(45$\degr$), 
i(90$\degr$)} and \textit{i(135$\degr$)}, obtained from the 
subimages taken in the polarimetric mode of the ISAAC instrument, has a 
\textit{FOV} of $\sim 100'' \times 200''$. In all cases it includes the 
entire region where the X-ray PWN is identified. The detected diffuse
emission, which 
can be seen in the total intensity image $I$ (Fig.~\ref{cfht-img}, \textit{bottom} panel), 
is restricted only to the very central region of the PWN. It is slightly 
extended and compatible in size and position with the emission blob 
detected with AOB-KIR in the \textit{K'} filter. 

Measurements of the degree of linear polarisation $P_{\mathrm{L}}$ and 
the polarisation angle $\theta$ reveal highly polarised emission from the nebula
with a regular structure of the electric field (Fig.~\ref{isaac-pol-vect}, panel \textit{b}). 
As can be seen from the $P_{\mathrm{L}}$ map (Fig.~\ref{isaac-pol-vect}, panel \textit{a})
this polarised emission extends within the X-ray compact nebula and is only 
important where the star contamination is minimal. 
To the NE of the pulsar position, but still within the X-ray compact nebula, 
a region where the linear polarisation degree is smaller than 0.1 can be identified 
(Fig.~\ref{isaac-pol-vect}, panel \textit{a}). As seen from the total intensity image 
$I$ (Fig.~\ref{cfht-img}, \textit{bottom} panel), this is also the region where strong 
stellar contamination is present. 
The linear polarisation fraction averaged over the emission region is high, 
$P_{\mathrm{L}}^{avg}\simeq 0.47 \pm 0.23$; moreover, in some regions  
within the polarisation nebula $P_{\mathrm{L}}$ 
is even as high as $0.6$. The high degree of linear polarisation points 
to the synchrotron nature of the observed radiation and a highly
ordered magnetic field structure.

As mentioned before, the measurements of $\theta$ across the compact nebula 
reveal a regular structure of the radiation electric field which
is illustrated as white vectors in the \textit{b} panel of Fig.~\ref{isaac-pol-vect}.
In the central part of the nebula the electric field vectors are oriented
in the NE-SW direction. In the outer regions of the nebula, vectors tend
to bend away from the NE-SW direction in a twofold way: moving to the NW vectors rotate
anticlockwise; in the SE part vectors rotate clockwise with respect to 
the electric field direction determined in the central part of the polarisation nebula. 
The change in orientation of the electric field vectors 
is accompanied by a steady decrease in the linear polarisation fraction $P_{\mathrm{L}}$.

A circular aperture similar in size to that used for the CFHT flux 
measurements was used to extract the counts from the detected polarised 
emission blob. Star contamination to the measured flux was reduced by 
their standard photometry measurements. 
The dereddened flux for the emission blob derived from the averaged 
total intensity is $F_{Ks} = (1.28 \pm 0.84)$~mJy; 
although it is compatible within uncertainties, 
we consider this value less reliable than the one obtained from
the AOB-KIR observations (see Sect.~\ref{spec}).

\begin{figure*}
  \begin{center}
    \begin{tabular}{cc}
	\includegraphics[trim= 10mm 25mm 0mm 5mm, clip, width=8cm]{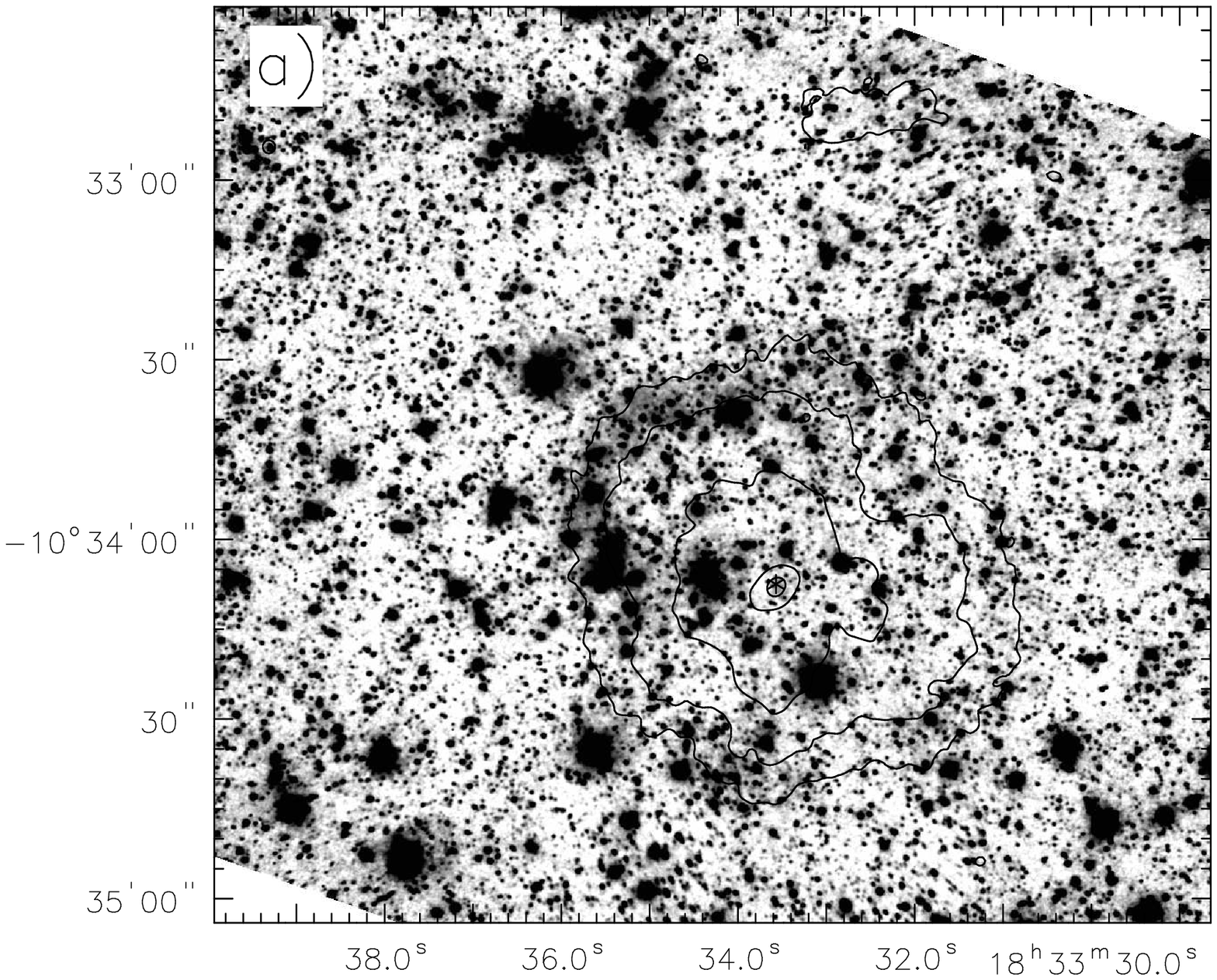}& 
	\includegraphics[trim= 10mm 25mm 0mm 5mm, clip, width=8cm]{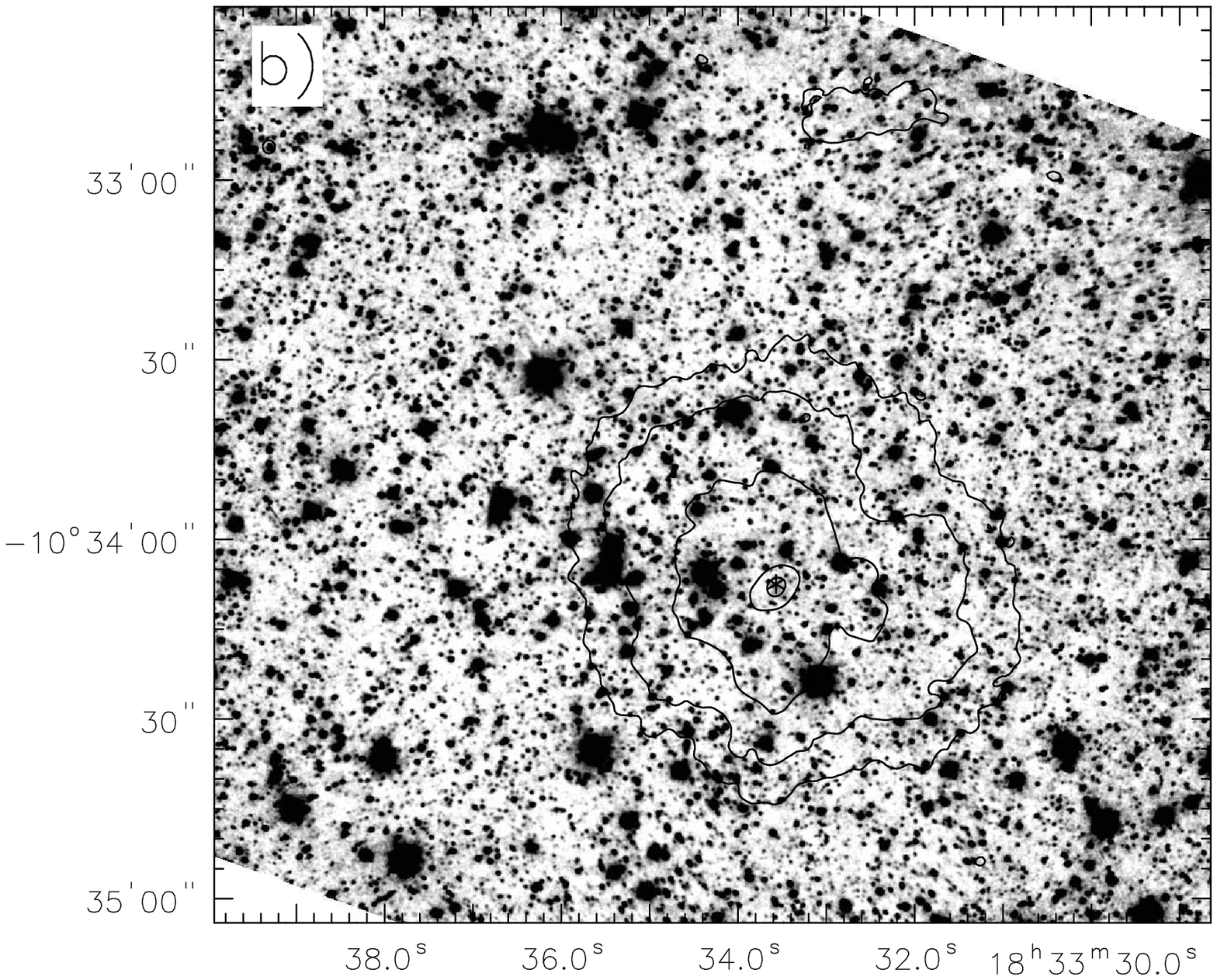}\\
	\multicolumn{2}{c}{\includegraphics[trim= 20mm 5mm 0mm 80mm, clip, width=12cm]{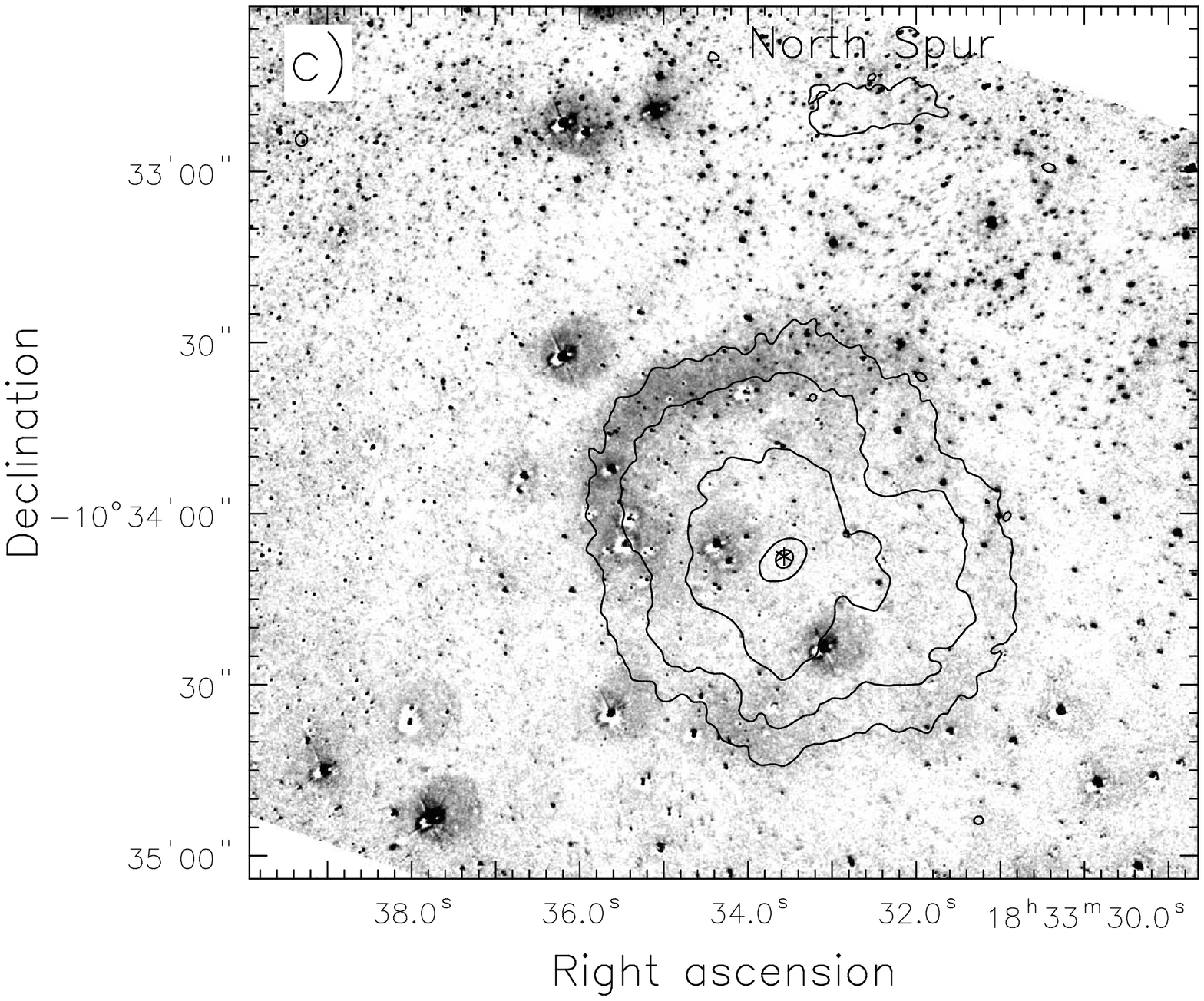}}\\
	\end{tabular}
	\caption{\small 	
	\textbf{a)} ISAAC mosaicked image of SNR G21.5-0.9 
	in the [Fe~II]~1.64~$\mu$m narrow-band filter. The full field of view is presented
	to show also the North Spur region. \textbf{b)} Same field 
	of view as in panel \emph{a}, but imaged in the 1.71~$\mu$m narrow-band 
	filter. Faint diffuse emission at the NE and SW of the PWN is visible in 
	1.64~$\mu$m filter, and not in 1.71~$\mu$m. 
	\textbf{c)} The [Fe~II] 1.64~$\mu$m image after subtraction
	of the image in the 1.71~$\mu$m narrow-band filter. 
	Some stellar residua are still visible;	the net flux in these residua 
	is typically around 9\% (RMS) of the corresponding stellar flux. 
	For all panels: the star depicts the position of PSR~J1833-1034; contours represent 
	the extent of the X-ray PWN seen by {\it Chandra} ACIS \citep{sln00}. 
	All images are presented with the same intensity scale. 
	For presentation purposes, images were smoothed with a 2-dimensional 
	Gaussian with $\sigma = 0.28''$.
	The images shown include the data from the first night
	of observation (08.07.2002) only, in order to yield more uniform
	coverage of the PWN and North Spur regions.}
	\label{isaac-img-fe2}
  \end{center}
\end{figure*}

\subsection{Compact Core -- mid-infrared}
\label{sub-sptz}

No large-scale extended emission associated with the PWN is seen in IRAC bands. 
However, in the PWN centre a compact emission blob is present (Fig.~\ref{sptz-f1}).  
It is clearly visible in all IRAC bands and its extent is comparable with the size 
of the putative pulsar wind torus seen in X-rays \citep{cam06}
and in the near-infrared.
For the IRAC observations the source flux was obtained
using an extraction region with a radius of $3.6\arcsec$ (large circle in the \textit{top left} 
panel of Fig.~\ref{sptz-f1}).
To estimate the background level, circular regions with a $2\arcsec$ radius (small circles 
in the top \textit{left panel} of Fig.~\ref{sptz-f1})
were chosen in the vicinity of the source extraction region. 
The average value from these regions was taken
to represent the background level at the source position.
Because the aperture used for the flux estimation in the IRAC bands was different
from the aperture used for the \textit{Spitzer} calibration stars an aperture 
correction, as described in IRAC Handbook\footnote{http://ssc.spitzer.caltech.edu/irac/iracinstrumenthandbook/}, 
was applied to the computed fluxes.
In order to estimate the compact nebula's flux at MIPS 24~$\mu$m standard aperture photometry
was used. A source extraction region with radius of $7\arcsec$ was used. The background 
was estimated using an annulus with an inner radius of $7\arcsec$ and an outer
radius of $13\arcsec$. To obtain the final value of the measured flux an aperture
correction was applied as described in the MIPS Handbook\footnote{http://ssc.spitzer.caltech.edu/mips/mipsinstrumenthandbook/}.
\textit{Spitzer} IRAC and MIPS 24~$\mu$m fluxes are presented in Table~\ref{sptz-t1}.
The flux uncertainties reflect the variations in the background level between different 
background regions,
as well as the uncertainty in the extinction correction for the
de-reddened fluxes.

\begin{table}
\caption{\textit{Spitzer} IRAC and MIPS observed and extinction corrected fluxes 
of the compact nebula. Correction for interstellar extinction was performed
using the local ISM table of \citet{ct06}, where $A_{\mathrm{K}} = 1.15$
was estimated from $N_{\rm H} = 2.2 \times 10^{22} {\rm cm}^{-2}$
following \citet{gor75} 
and \citet{crdl89}.}
	\begin{center}
	\begin{tabular}{cccc}
	\hline\hline
	Instrument & $\lambda$ & Observed Flux & De-reddened Flux\\
	 & [$\mu$m] & [mJy] & [mJy]\\\hline
	 IRAC &  & &\\
	  & 3.6 & 0.53 $\pm$ 0.19 & 0.94 $\pm$ 0.45 \\
	  & 4.5 & 0.74 $\pm$ 0.14 & 1.19 $\pm$ 0.49 \\
	  & 5.8 & 0.88 $\pm$ 0.20 & 1.34 $\pm$ 0.49 \\
	  & 8.0 & 0.89 $\pm$ 0.22 & 1.38 $\pm$ 0.71 \\
	 MIPS & & &\\	 
	  & 24.0 & 2.47 $\pm$ 0.67 & 4.37 $\pm$ 2.13 \\
	  \hline
	\end{tabular}
	\end{center}
	\label{sptz-t1}
\end{table}

\subsection{Pulsar Wind Nebula}
\label{sub-pwn}

The mosaicked image taken in the [Fe~II] 1.64~$\mu$m and 1.71~$\mu$m 
filters (see Sect.~\ref{sub-img}) covers a \textit{FOV} of $\sim 230'' 
\times 160''$. This includes both the PWN region and the North Spur 
region (a bright spot located in the northern part of the X-ray halo 
surrounding the PWN \citep{bocc05}). 
By subtracting the 1.71~$\mu$m image (corrected for the PSF and average
star flux difference between the filters) from the [Fe~II] 1.64~$\mu$m image 
(see Sect.~\ref{sub-img}) we minimise the stellar 
contamination to any [Fe~II] 1.64~$\mu$m line emission that can be present 
in the observed region. The resultant image is presented in panel \textit{c}
of Fig.~\ref{isaac-img-fe2}. No [Fe~II] line emission associated with the North Spur is detected.
There is, however, extended ($\sim 80''$ in diameter) [Fe~II] emission associated with the PWN. 
Its shape follows the outer contours of the X-ray nebula (Fig.~\ref{isaac-img-fe2},
panel \textit{b}). 
The brightness distribution is non-uniform; a partial ring-like structure has a maximum 
of emission in the NE part. Going towards the SW, the emission fades leaving only a few faint 
patches of [Fe~II] emission which fall within the X-ray contours of the PWN (Fig.~\ref{isaac-img-fe2}, 
panel \textit{c}). The brightest NE portion of the ring shows clear limb-brightening with a well-defined 
outer edge.
To the SW from the pulsar position an emission blob, probably an artefact
associated with the bright star (as seen next to other bright stars in
the FOV) rather than with the PWN material, is apparent. 
The mean surface brightness of the NE part of [Fe~II] line emission is
$S_{\mathrm{[Fe~II]} 1.64\mu\mathrm{m}} = (3.3 \pm 0.5)\times10^{-16}$ erg s$^{-1}$ cm$^{-2}$
arcsec$^{-2}$ while for the southern rim it is $S_{\mathrm{[Fe~II]} 1.64\mu\mathrm{m}}
 = (1.7 \pm 0.3)\times10^{-16}$ erg s$^{-1}$ cm$^{-2}$ arcsec$^{-2}$.

\section{Interpretation}
\label{interp}

\begin{figure}
  \begin{center}
	\includegraphics[scale=0.5]{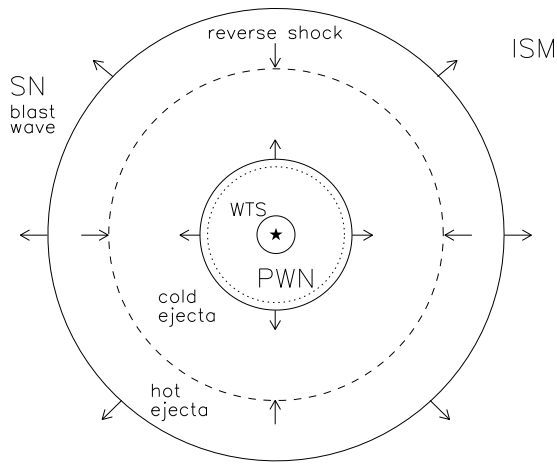}
	\caption{\small Schematic view of a young composite supernova remnant structure
	(not to scale). 
	The supernova (SN) blast wave propagates in the interstellar medium (ISM). 
	Inside the SN remnant a pulsar wind nebula (PWN) powered by the pulsar 
	(star) is formed. The PWN expands into the cold SN ejecta, which from 
	the outer side are being heated/compressed by the reverse shock facing 
	inwards in the SN remnant. Inside the PWN, where the pulsar wind pressure 
	balances the pressure of the surrounding matter a wind 
	termination shock (WTS) is formed. The inner solid line indicates the shock 
	driven into the inner ejecta by the accelerating PWN. Dotted line depicts 
	the position of the contact discontinuity separating the shocked SN ejecta 
	overtaken by the PWN outer shock, and the shocked pulsar wind plasma having 
	gone through the WTS. Detailed discussion of SNR/PWN structure can be found 
	in e.g.	\citet{kc84} or \citet{tmck99}.}
	\label{snr-scheme}
  \end{center}
\end{figure}

Both the emission detected with AOB-KIR, \textit{Spitzer} and the ISAAC polarimetric 
observations and also the structure seen in the [Fe~II] 1.64~$\mu$m emission 
line, can be interpreted within a composite supernova remnant evolutionary 
scheme \citep[e.g.][]{gaenslane06, vds01, rc84}. According to this 
evolutionary scheme a supernova (SN) explosion drives a blast wave into 
the surrounding interstellar medium (ISM). At the same time in the 
interior of the expanding SNR a pulsar wind nebula evolves. It is powered 
by the newly born pulsar through its relativistic magneto-hydrodynamic wind. 
At early times, the pulsar energy input causes the PWN's expansion 
to accelerate, driving a shock into the inner edge of the uniformly expanding 
ejecta.
Inside the PWN, at the point where the pulsar wind ram pressure is 
equilibrated by the pressure exerted by the surrounding medium, a pulsar 
wind termination shock (WTS) is formed. The schematic structure of 
such a young composite SNR is presented in Figure~\ref{snr-scheme}.

\subsection{Polarisation - Wind Termination Shock}
\label{wts}
The high degree of linear polarisation measured across the emission blob 
seen in the \textit{Ks} band (see Sect.~\ref{sub-cc}) points to its 
synchrotron origin. Simultaneously, the orientation of the electric 
field vectors (Fig.~\ref{isaac-pol-vect}, \textit{bottom} panel) suggests 
that the region where synchrotron radiation is produced has a toroidal 
magnetic field. Such conditions are expected to be found in the wind 
termination shock region (Fig.~\ref{snr-scheme}). The relativistic particles 
flowing upstream of the shock do not radiate. Only after being accelerated 
at the shock, can they produce synchrotron radiation in the downstream flow. 
The best example in support of such a scenario is the torus in the Crab Nebula
\citep[][and references therein]{hest08}.
When viewed in polarised light it shows magnetic field structure that is 
aligned along wisps and fibrous structures which are present in the torus,
and which follow its shape. For some of the wisps a high degree of linear 
polarisation, near 0.7, is measured.

Recent MHD simulations studying optical polarisation in the inner parts
of pulsar wind nebulae \citep{bcc05,delz06,vol09} predict that emission from
a wind termination shock region should be highly polarised. In these simulations
a pure toroidal structure of magnetic field and a radial flow of particles
are assumed. The maximum value of linear polarisation fraction, close to 70\%, 
is expected from the WTS region close to the torus symmetry axis as projected on the sky. 
When moving away from the axis, the polarisation progressively decreases to values 
smaller than 50\%. In addition, close to the torus edges, regions that are almost 
completely depolarised can be found \citep[see Fig.~2 of][]{bcc05}. 
Simulations also show that the polarisation angle should change across the emitting 
region. In the parts close to the symmetry axis of the torus, the electric vector
should be aligned with the axis and its inclination with respect to the axis
should increase while moving away from it. The region where the electric
vector becomes perpendicular to the symmetry axis depends on the inclination
angle of the torus with respect to the sky, and also on the bulk flow speed. Concerning
the latter dependence, the position where the polarisation vector becomes
perpendicular to the axis moves closer towards the axis with increasing
value of the bulk speed \citep{bcc05}.

Similar behaviour to that predicted by MHD simulations is found in our
polarisation observations of the central part of G21.5-0.9 (see Sect.~\ref{sub-cc}),
which supports the claim that in the region of the wind termination shock a toroidal
magnetic field is present. From the investigation of the observed polarisation angle 
swing, the probable orientation of the symmetry axis of the toroidal field as 
projected on the sky is $\Psi \sim 40\degr$ (measured North to East). In this 
scenario, $\Psi$ is also the position angle of the pulsar spin axis as projected on 
the sky, which is in agreement to the value proposed by \citet{cam06}. 
The nebula seen in the \textit{Ks} band can be described as an ellipse 
with semimajor axis of $\theta_{1} \simeq 2.9''$ and semiminor axis of 
$\theta_{2} \simeq 1.8''$. The ellipse minor axis is oriented at $\Psi$, so
it is parallel to the toroidal magnetic field symmetry axis as projected on the sky. 
However, due to the Doppler boosting experienced by the relativistic particles 
escaping the wind termination shock, the visible size of the emission region 
cannot be simply translated into the shock radius. 
A vital part of the WTS may be invisible because the size of 
the emitting region highly depends on the bulk flow speed 
in the WTS region \citep{bcc05}. 
If the emitting torus is not seen edge-on, but is inclined with respect to 
the plane of the sky, the Doppler boosting effect could possibly explain 
the shift of the compact nebula emission towards SW with respect to 
the pulsar position. To more accurately estimate the orientation of the toroidal 
magnetic field and the flow speed, detailed modelling is needed. However,
by simple comparison with the numerical results of \citet{bcc05}, \citet{delz06} 
and \citet{vol09} one may expect that the particle flow speed in the WTS region 
is high ($>0.2c$).

\subsection{Nonthermal Emission Spectrum}
\label{spec}

\begin{figure}
  \begin{center}
	\includegraphics[scale=0.5]{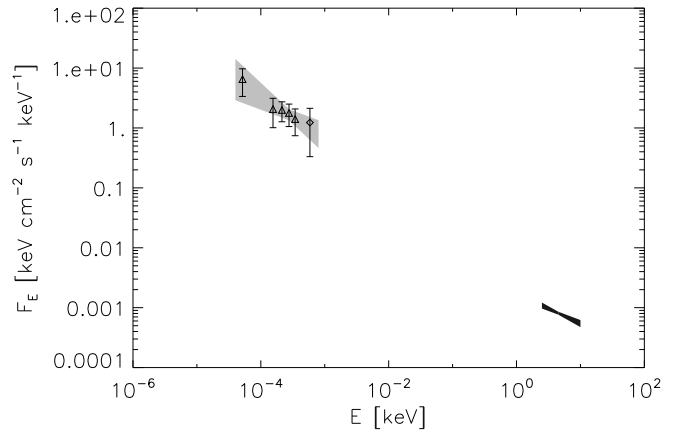}
	\caption{\small Spectral energy distribution of the compact nebula.
	The diamond shows the flux obtained with CFHT/AOB-KIR. \textit{Spitzer} 
	measurements obtained in the IRAC 3.6, 4.5, 5.6, 8.0~$\mu$m bands 
	and the MIPS 24~$\mu$m band are presented with triangles. All the infrared 
	fluxes are corrected for interstellar extinction. Light grey bow tie 
	region shows the uncertainty of the power-law fit to the infrared fluxes. 
	The fitted IR photon index is $\Gamma_{\mathrm{IR}} = 1.7 \pm 0.3$. The black
	bow tie represents the intrinsic X-ray spectrum (with its uncertainty) 
	of the compact nebula as obtained from {\it Chandra} observations
	\citep{cam06}. It is described by the photon index $\Gamma=1.5\pm0.1$ 
	(see text).
	}
	\label{sed-cc}
  \end{center}
\end{figure}

Figure~\ref{sed-cc} presents the spectral energy distribution of the compact 
nebula based on our flux measurements 
in the CFHT/AOB-KIR, \textit{Spitzer} IRAC 3.6, 4.5, 5.6, 8.0~$\mu$m and MIPS 24~$\mu$m 
bands, and its intrinsic X-ray spectrum obtained from {\it Chandra} observations presented 
by \citet{cam06}. The plotted IR fluxes are the values corrected for the interstellar 
extinction using the local ISM table of \citet{ct06}, where $A_{\mathrm{K}} = 1.15$
was estimated from $N_{\rm H} = 2.2 \times 10^{22} {\rm cm}^{-2}$
following \citet{gor75} and \citet{crdl89}. A simple power law $F_{\mathrm{IR}} = A 
E^{-\alpha_{\mathrm{IR}}}$, where $\alpha_{\mathrm{IR}}$ 
is the spectral index, was fitted to the infrared fluxes using a least-squares 
fitting method that takes into account the uncertainties of the measured fluxes as 
their weights. 
In the fit the flux value obtained 
from the VLT/ISAAC-POL averaged total intensity $I$ was not included because the flux value from
CFHT/AOB-KIR observations is a more accurate representation of the compact nebula flux
at $\sim 2.1\mu$m. This is because the higher resolution (smaller PSF) of CFHT/AOB-KIR
allows for much more accurate stellar subtraction in the crowded compact nebula 
field, which is difficult in the case of VLT/ISAAC-POL and in turn leads 
to an overestimation of the measured flux. 
The fitting to the rest of the IR fluxes yields the spectral index 
$\alpha_{\mathrm{IR}} = 0.7 \pm 0.3$. The obtained power law with its uncertainty 
is presented in Fig.~\ref{sed-cc} as the light grey bow tie region.

The X-ray spectrum in the $0.5-10$~keV energy range was obtained from 
{\it Chandra} ACIS observations \citep{cam06} using a circular extraction region 
with $4\arcsec$ radius around the PSR. The background level was estimated using 
an annulus with radius $4\arcsec - 10\arcsec$. 
Using the best-fit value of $N_{\mathrm{H}} = 2.2 \times
10^{22}$~cm$^{-2}$ for the entire nebula \citep{sln00} this yields
a photon index of $\Gamma = 1.46\pm 0.02$ and unabsorbed flux 
in the $2.5-10$~keV energy range 
$F_{2.5-10\mathrm{keV}} = (7.8 \pm 0.2) \times
10^{-12}$~erg~s$^{-1}$~cm$^{-2}$. Allowing the column density to
be determined just from the fit of the compact region yields similar
values of $N_{\mathrm{H}} = (2.3 \pm 0.1) \times 10^{22}$~cm$^{-2}$,
$\Gamma = 1.56 \pm 0.04$ and $F_{2.5-10\mathrm{keV}} = 7.7^{+0.5}_{-0.1}
\times 10^{-12}$~erg~s$^{-1}$~cm$^{-2}$.  
Comparison of
the fitted IR spectrum with the X-rays (Fig.~\ref{sed-cc}) suggests
that the spectrum must flatten between infrared and
X-rays.  Given the unusual implications of such a result on the
underlying particle spectrum, it is important to consider possible
factors that could complicate the interpretation of the X-ray
spectrum. There are two primary sources of potential bias: contamination
from the central pulsar, and dust scattering effects.

Regarding the pulsar contamination, to estimate a point source
contribution to the unabsorbed flux of the compact nebula, \citet{cam06} 
use {\it Chandra} HRC data to model the spatial emission around
PSR J1833-1034 as a point source and a 2-D Gaussian.  Converting their
best-fit count rate of the point-like component into a flux using
$N_H = 2.2\times10^{22}$~cm$^{-2}$ and an assumed photon index of
1.5 yields an unabsorbed flux $F_{2.5-10\mathrm{keV}}^{\mathrm{PSR}}
= 1.8\times 10^{-13}$~erg~s$^{-1}$~cm$^{-2}$ (about 2\% of the
unabsorbed flux obtained for the compact nebula).  This contribution
is insufficient to account for the apparent difference in spectral
index between the X-ray and IR bands (and the reduced flux upon
subtracting the point source contribution marginally increases the
difficulty in matching the spectra).

The properties of X-ray scattering from dust grains are well known 
\citep[see e.g.][]{sd98}, but modelling of X-ray halos of diffuse sources 
is a difficult task, because the effect is non-local. 
\citet{bocc05} have modelled the X-ray halo detected around G21.5-0.9, 
showing that it is mostly an effect of dust scattering, deriving halo 
parameters, like the scattering optical depth $\tau_{\mathrm{scatt}} 
\simeq 0.8 E_{\mathrm{keV}}^{-2}$: from this energy dependence it 
is apparent that lower-energy photons are scattered with much higher 
efficiency. An additional effect of scattering is a radial 
dependence of the best-fit column density ($N_{\mathrm{H}}$), which 
in G21.5-0.9 ranges from $\sim 1.5\times 10^{22}$~cm$^{-2}$ in the outskirts
of the PWN, up to $\sim 2.5\times 10^{22}$~cm$^{-2}$ in the very central 
regions \citep[see Fig.~3 in][]{bocc05}.

A consequence of scattering is a deficiency of soft photons in the
central regions, which leads standard spectral analysis routines
to overestimate the absorption column density; conversely, the
surrounding regions receive the scattered photons and show an excess
of soft photons that leads to a negative bias on the estimated
$N_{\mathrm{H}}$.
Since in principle similar biases could affect the estimates of the spectral 
index and unabsorbed flux, depending on the details of the data analysis, 
we set upper limits to those possible biases. 
We have chosen the energy range $2.5-10$~keV for our fits to the
spectrum of the inner torus: this is a good compromise for weighting
harder photons (only weakly affected by scattering) but still
retaining good statistics. With this energy cut we cannot independently
determine the value of $N_{\mathrm{H}}$, but instead consider a
range of $N_{\mathrm{H}}$ parameters, chosen from 2.0 to
$2.4\times10^{22}$~cm$^{-2}$ ($2.2\times10^{22}$~cm$^{-2}$ being
the average value obtained from a fit to the whole PWN).

Another question is whether or not to use as background the emission of 
the surrounding PWN: the emission in the surroundings of the torus is partly 
contaminated by the halo of the torus itself, 
and some of the photons emitted by the outer PWN are scattered
into the area on which the torus extends.
In Table~\ref{gam-est} we present the results for the grid 
of cases mentioned above. From this we can conclude that conservative
error intervals
on the parameters, taking also into account dust scattering effects, 
are the compact nebula photon index of $\Gamma = 1.5\pm 0.1$ and the 
unabsorbed flux of $F_{2.5-10\mathrm{keV}} = (8.7 \pm 1.0) \times 
10^{-12}$~erg~s$^{-1}$~cm$^{-2}$.

\renewcommand\arraystretch{1.5}
\begin{table}
\caption{Grid of hydrogen column densities $N_{\mathrm{H}}$ used for determination of the X-ray spectrum 
		and the unabsorbed flux in the $2.5-10$~keV range. Fits were performed using an annulus with radius 
		of $4\arcsec-10\arcsec$	for background level estimation (\emph{top} part of the table), and not using 
		any background region (\emph{bottom} part of the table).}
 \begin{center}
	\begin{tabular}{ccc}
	\hline\hline
	N$_{\mathrm{H}}$  &  $\Gamma$  &  $F_{2.5-10\mathrm{keV}}$\cr
	[$10^{22}$~cm$^{-2}$] & & [$10^{-12}$~erg~s$^{-1}$~cm$^{-2}$]\\[0.5ex] \hline
    \multicolumn{3}{c}{Background from surrounding annulus}\\[0.2ex] \hline
	2.0 &   1.43 $^{~+~0.04}_{~-~0.05}$  &     7.7 $^{~+~0.2}_{~-~0.3}$ \\
	2.2 &   1.47 $^{~+~0.05}_{~-~0.04}$  &     7.6 $^{~+~0.2}_{~-~0.2}$ \\
	2.4 &   1.51 $^{~+~0.05}_{~-~0.05}$   &     7.7 $^{~+~0.2}_{~-~0.2}$ \\
	&&\cr \hline

	\multicolumn{3}{c}{No background}\\[0.2ex] \hline
	2.0 &   1.49 $^{~+~0.04}_{~-~0.04}$  &     9.6 $^{~+~0.2}_{~-~0.2}$ \\
	2.2 &   1.53 $^{~+~0.04}_{~-~0.03}$  &     9.7 $^{~+~0.2}_{~-~0.2}$ \\
	2.4 &   1.57 $^{~+~0.04}_{~-~0.03}$  &     9.7 $^{~+~0.2}_{~-~0.2}$ \\
	\hline
	\end{tabular}
 \end{center}
 \label{gam-est}
\end{table}

We note that a joint fit between the X-ray and IR data, leaving all
parameters free, yields $N_H = (2.6 \pm 0.2) \times10^{22}$~cm$^{-2}$
and $\Gamma = 1.76^{+0.07}_{-0.05}$. Applying these values to the
X-ray data alone provides a good fit ($\chi^2_r = 1.1$), though the
X-ray fit obtained with $N_H$ and $\Gamma$ as free parameters is
statistically much better ($\chi^2_r = 1.01$). The results are similar
when a dust-scattering component is included in the fitting process.
While the spectral modeling thus argues for a spectral flattening
between the IR and X-ray bands, it is clearly important to search
for examples of such behavior in other sources given that the results
here rely exclusively on somewhat modest differences in parameters
from spectral fitting.

Even though the torus in G21.5-0.9 is not detected in the radio, it would 
appear that a further break or steepening must exist between the
radio and IR bands.
Using the radio map at 5~GHz \citep{bb08} we can estimate an upper limit 
$F_{5\mathrm{GHz}}^{\mathrm{UL}} \simeq 52$~mJy to the radio flux of the compact 
nebula by integrating the signal within a radius of $3\arcsec$ around the pulsar 
position.
This can be compared to the 5~GHz flux $F_{\mathrm{ext}} 
\simeq 1.1$~Jy which one gets by extrapolating the fitted IR spectrum to the radio band. 
This extrapolated value $F_{\mathrm{ext}}$ is much higher than the
estimated upper limit $F_{5\mathrm{GHz}}^{\mathrm{UL}}$, which argues in
favour of the existence of spectral break between the radio and IR bands.
This would be expected because the radio spectrum of the torus presumably 
has the same index as the entire PWN (whose radio index is flatter than what 
has been measured here for the torus in the IR), because the particles responsible 
for the radio emission will not have suffered significant radiative losses 
in filling the nebula.
Similar spectral behaviour is also observed for the torus in 3C~58 \citep{sln08}.
These two cases show that the injection spectrum is complex, and not a pure power law.
\citet{fb07} have calculated synchrotron spectra of a power-law distribution of electrons
radiating in magnetic turbulence of prescribed properties. For appropriate choices of 
parameters, it is possible to produce a spectrum that could continue to steepen from 
radio to X-ray energies, as our data imply; however, the suggestion in our data of 
a flattening of the spectrum between IR and X-rays cannot be produced by such a model.
Moreover, the \citet{fb07} calculations were applied to the integrated spectra
of the PWNe and as such they may not be applicable to explain the spectra produced 
in the vicinity of the wind termination shock, which is the case for the spectrum 
presented in Fig.~\ref{sed-cc}.

\subsection{[Fe~II] filaments - PWN shock}
\label{fe2}
Studies carried out in the past 20 years on infrared emission from shell 
supernova remnants show that [Fe~II] emission seen in shell 
SNRs is either induced by the reverse shock propagating back through 
the fast moving ejecta or produced at the forward shock from iron 
present in the circumstellar and/or interstellar medium.
In the case of pulsar wind nebulae, by contrast, iron emission, if collisionally 
excited, is produced by the rapidly expanding PWN encountering the slow-moving 
cold ejecta (Fig.~\ref{snr-scheme}).
This interpretation has been applied to observed [Fe~II] lines at 17.9 and 26~$\mu$m 
from G54.1+0.3 \citep{tm10} and B0540-69.3 \citep{wil08}. Our imaging observations 
demonstrate that in SNR G21.5-0.9, the
iron emission does in fact originate at the location 
of the PWN-ejecta interaction. The nature of this emission then characterizes 
the otherwise unobservable innermost supernova ejecta, material formed just above 
the mass cut within which material fell back to become the neutron star.

In G21.5-0.9 the detected [Fe~II] 1.64~$\mu$m emission - having a ring-like 
structure that is significantly brighter in the NE - follows the outer edge 
of the PWN (Fig.~\ref{isaac-img-fe2}). 
This behaviour suggests that the [Fe~II] emission 
is caused by the PWN driven shock that compresses the surrounding cold 
supernova ejecta (Fig.~\ref{snr-scheme}). Collisional excitation would
then take place in the medium just behind the shock front. 
The low excitation energy means that only low shock velocities 
are required, perhaps as a result of shocks being driven into dense clumps 
in the ejecta, but it is also possible that the PWN forward-shock speed 
is slow due to its expansion into moving ejecta. Figuring out which one is more 
plausible in the case of G21.5-0.9 would require a dynamical model and/or spectroscopic 
study in search of density diagnostics in the line-emitting regions.
The observed brightness difference between regions in the NE and SW could 
result from a density difference in the medium that the expanding PWN 
bubble encounters, which in turn would imply a difference in the PWN shock 
velocity in the NE part with respect to the SW. 

A good example of the PWN expanding into a medium of nonuniform
density is the Crab Nebula. Such conclusion is drawn based on observations
of [O~III] and [Ne~V] emission, excited by the PWN shock, which 
forms so-called skin at the boundary of the Crab Nebula. The skin
is visible around most of the outer edge of the nebula with
the exception of its NW part \citep{sanhest97}. The lack of [O~III] 
and [Ne~V] emission is explained as due to variations in the density
of the ejecta into which the nebula is expanding. This view is also 
supported by the morphology of Rayleigh-Taylor fingers \citep{loll07}, which shows SE-NW
asymmetry. In the NW part of the Crab Nebula there are fewer fingers
comparing to the SE. This asymmetry results from a preshock ejecta density
lower in the NW than in the SE \citep[][and references therein]{hest08}.

However, another possible scenario may explain the [Fe~II] 
1.64~$\mu$m line emission in PWNe. \citet{grah89} showed 
that the near-infrared iron emission detected in the Crab Nebula 
originates from optically thick filaments which are photoionized by 
the UV--X-ray component of the synchrotron continuum present in 
the PWN.
In the general ISM, a measured line ratio ([Fe~II]~1.64~$\mu$m)/(Br$\gamma$)~$\gg$~1
is interpreted as the case of shock excitation, as radiative shocks destroy grains 
and return Fe to the gaseous state. In SNR ejecta, however, we do not know what 
fraction of Fe might be in grains initially, or what the Fe/H abundance ratio is, making 
this test inappropriate. In addition, as pointed out by \citet{grah89}, in the Crab Nebula 
conditions leading to high values of ([Fe~II]~1.64~$\mu$m)/(Br$\gamma$) can also 
be found in the photoionized filaments.
Thus, determining this line ratio 
as the sole factor discriminating between the two scenarios may not
be conclusive. Recent studies of SNRs where [Fe~II] 1.64~$\mu$m is detected,
e.g. G11.2-0.3 \citep{koo07} and 3C~396 \citep{lee09}, show that iron emission 
can be explained by a radiative shock propagating into the gas with a velocity 
$\sim 100$~km~s$^{-1}$. The iron line-emitting regions are characterised by 
high electron densities $n_{\mathrm{e}} \sim 10^{4}$~cm$^{-3}$ and electron
temperatures $T_{\mathrm{e}}$ of order a few times 10$^{3}$~K. This can be
contrasted with the conditions found for the photoionized case in the Crab Nebula 
where $n_{\mathrm{e}} \simeq 200$~cm$^{-3}$ and $4000 < T_{\mathrm{e}} < 8000$~K
\citep{grah89}. Thus in order to try to discriminate between the two scenarios
in the case of G21.5-0.9, spectroscopic observations of the [Fe~II] line-emitting 
regions are needed.

\section{Conclusions}
\label{concl}

In this paper we report the first detection in the infrared band
of emission emanating from the compact nebula region in SNR~G21.5-0.9 
($\sim$2$\arcsec$ radius around the pulsar position) through imaging and polarimetry. 
The toroidal structure of the magnetic field in the compact nebula is 
inferred from polarimetric observations in the \textit{Ks} band. The high
degree of linear polarisation together with the observed
swing of the electric field vector across the nebula confirms
that the observed radiation is synchrotron emission from relativistic
particles accelerated at the wind termination shock and radiating in
the downstream flow.
The infrared spectrum of the compact nebula can be described as a
power law with spectral index $\alpha_{\rm IR} = 0.7 \pm 0.3$; when
combined with X-ray results from {\it Chandra} observations \citep{cam06},
it suggests that the spectrum flattens between the infrared and
X-ray bands.

The detection of [Fe~II] 1.64~$\mu$m emitting material within the PWN of G21.5-0.9
is also reported. This iron emission is detected in the NE and in the SW of 
the PWN. The determined surface brightness of the SW emission is weaker 
compared to that of the NE region. It is important to
point out that the detected [Fe~II] emission has the shape of a limb-brightened 
broken ring ($\sim$40$\arcsec$ radius), which follows the edge of the X-ray 
and radio PWN, but is concentrated inside the PWN boundary. Most probably the observed iron 
emission originates from cold supernova ejecta shocked by the expanding PWN bubble. 
In this case collisionally excited iron radiates in the medium 
behind the shock. However, to confidently discriminate between the case 
of shock excitation and photoionization by the synchrotron continuum, spectroscopic 
observations of the iron line-emitting material are needed.
Line profiles can give shock velocities, while observation of additional 
spectral lines can provide density and temperature diagnostics to constrain shock 
models. Important information to be gained this way can include the speed of shocks 
causing excitation and ionization; the expansion rate of the PWN; abundances of the shocked material; 
and possible contributions of dust to the IR continuum, allowing inferences on newly synthesized dust.
G21.5-0.9 thus joins the small elite class of objects, along with G54.1+0.3 and B0540-69.3, 
where IR observations of PWN interactions are illuminating the deep interior of 
a core-collapse supernova.
 
The present work demonstrates that infrared observations, both imaging and polarimetry,
are an effective tool for detecting optically obscured pulsar wind nebulae 
and for studying their inner structure.

\begin{acknowledgements}
\label{acknow}
This research is based in part on observations obtained at the Canada-France-Hawaii 
Telescope (CFHT) which is operated by the National Research Council of Canada, 
the Institut National des Sciences de l'Univers of the Centre National de 
la Recherche Scientifique of France, and the University of Hawaii.
It is also based on observations made with ESO Telescopes at the La Silla 
Paranal Observatory under programme ID 69.D-0279(A, B).
This work is based in part on observations made with the \textit{Spitzer} Space 
Telescope, which is operated by the Jet Propulsion Laboratory, California 
Institute of Technology under a contract with NASA.
This publication makes use of data products from the Two Micron All Sky Survey, 
which is a joint project of the University of Massachusetts and the Infrared 
Processing and Analysis Center/California Institute of Technology, funded 
by the National Aeronautics and Space Administration and the National 
Science Foundation.
We thank M.~F. Bietenholz for providing the radio map of G21.5-0.9 in
numerical form, and an anonymous referee for comments which substantially 
improved the clarity of the paper.
This research was partially supported by MNiSW grant N203 387737 and
\mbox{ECO-NET} grant 18874~WA. 
This work was carried out within the framework of the European Associated Laboratory
"Astrophysics Poland-France". 
A.~Zajczyk acknowledges partial support from the Scholarship 
for PhD students "Stypendium dla doktorant\'{o}w 2008/2009 - ZPORR". 
P.~Slane acknowledges partial support
from NASA contract NAS8-03060 and \textit{Spitzer} grant JPL 1265776.
\end{acknowledgements}

\bibliographystyle{aa} 	
\bibliography{zajczyk_etal_g21-infrared}

\end{document}